\documentclass[twocolumn]{aastex701}

\usepackage{amsmath}
\usepackage{multirow}

\newcommand{\fb}{f_{\rm{b}}}
\newcommand{\rl}{r_{\rm{Lag}}}
\newcommand{\wk}{W_0}

\newcommand{\msun}{M_{\odot}}

\newcommand{\pc}{\rm{pc}}
\newcommand{\kpc}{\rm{Kpc}}

\newcommand{\myr}{\rm{Myr}}
\newcommand{\gyr}{\rm{Gyr}}

\newcommand{\rhm}{r_{\rm{hm}}}
\newcommand{\rtid}{r_{\rm{tid}}}

\newcommand{\rv}{r_{\rm{v}}}
\newcommand{\cmc}{\texttt{CMC}}
\newcommand{\cosmic}{\texttt{COSMIC}}

\newcommand{\tpert}{t_{\rm{pert}}}

\newcommand{\trhm}{t_{\rm{r,hm}}}

\shorttitle{An equilibrium density profile for star clusters}
\shortauthors{Khurana \& Cen}

\begin{document}

\title{On the Existence of an Equilibrium Density Profile for Massive Collisional Star Clusters}

\author[orcid=0000-0003-3259-6702,gname=Ambreesh,sname=Khurana]{Ambreesh Khurana}
\affiliation{Center for Cosmology and Computational Astrophysics (C3A), Institute for Advanced Study in Physics, Zhejiang University, Hangzhou, China}
\affiliation{Institute of Astronomy, School of Physics, Zhejiang University, Hangzhou, China}
\email[show]{akhurana@zju.edu.cn}

\author[orcid=0000-0001-8531-9536,gname=Renyue,sname=Cen]{Renyue Cen}
\affiliation{Center for Cosmology and Computational Astrophysics (C3A), Institute for Advanced Study in Physics, Zhejiang University, Hangzhou, China}
\affiliation{Institute of Astronomy, School of Physics, Zhejiang University, Hangzhou, China}
\email[show]{renyuecen@zju.edu.cn}

\correspondingauthor{Ambreesh Khurana, Renyue Cen}

\begin{abstract}

The dynamical evolution of massive, dense star clusters is governed by a complex interplay of multiple physical processes, including two-body relaxation, strong encounters, stellar evolution, and galactic tides, superimposed on the orbital motion of individual stars and binaries within the cluster potential. These processes can alter the internal evolution of individual stars and binaries, produce exotic stellar objects, and collectively determine the global structural evolution of the cluster. In this paper, we use the Monte Carlo star cluster evolution code $\cmc-\cosmic$ to study how the density profiles of dense star clusters evolve under the combined effects of these processes. We find that for a given initial mass and binding energy, cluster models converge toward a characteristic density profile largely independent of their initial density distributions. This convergence occurs during the late-time, quasi-static phase of evolution, when gravitational contraction is balanced by energy generated through black hole binaries in the cluster core. The shape of this profile is well represented by a King profile with $\wk=6.8$ but with the virial radius of the final cluster larger than the initial by a factor of $\sim7$. We further show that this equilibrium profile is dynamically stable: perturbations away from it decay through the cluster’s natural dynamical evolution, driving the system back toward the equilibrium profile.

\end{abstract}


\section{Introduction} 
\label{sec:intro}

Massive and dense star clusters, such as globular clusters (GCs) or young massive clusters (YMCs), are among the most versatile astronomical targets in terms of their potential to uncover important results for almost all branches of astronomy and astrophysics. Composed of stars of similar age and metallicity, they are among the ideal targets for elucidating stellar evolution. Their rich internal dynamics makes them efficient in producing exotic astrophysical objects such as X-ray binaries, blue stragglers, cataclysmic variables, upper mass-gap black holes (BHs), and merging BH binaries \citep[e.g.,][]{Meylan_1997,PortegiesZwart_2010,Rodriguez2015,Khurana_2023,Khurana_2025,Sedda_2026}. GCs, specifically, also play an important role in extragalactic astronomy. As they are thought to have formed during the major merger events of their host galaxy \citep[e.g.,][]{Kruijssen2015,El-Badry2019a,Newton_2025}, they hold information about its assembly history \citep[e.g.,][]{Brodie_2006,Massari_2023,Valenzuela_2024}. In addition, GCs have been used to constrain the dark matter distribution of their host galaxy as well \citep[e.g.,][]{Valenzuela_2023,Reina-Campos_2023,Liang_2024}.

However, the very aspects that make them such interesting sources to study are also the reason they are challenging to study. E.g., for GCs, the combination of high density, large mass, and old age puts them in a parameter-space region where their age is typically larger than their relaxation time by a factor of a few \citep[e.g.,][]{HeggieHut_2003}, while still being massive enough that their computational modeling remains a challenge. This is in contrast to smaller open clusters or galaxies, where one of those two conditions is not satisfied. This means that a long and complicated evolution separates present-day GCs and their high redshift progenitors \citep[e.g.,][]{Brodie_2006,Chatterjee2013}.

One of the many interesting questions that we can ask about massive and dense star clusters is: what are their structural properties and how do they evolve with time? Several decades of work has been done to elucidate this question. 

Observationally, Milky Way GCs seem to be generally well described by King \citep[][]{King_1966} or Wilson \citep[][]{Wilson_1975} profiles \citep[e.g.,][]{Madsen_1996,McLaughlin_2005,Miocchi_2013}, except possibly the outskirts \citep[e.g.,][]{DeBoer_2019}. King profiles are motivated by the assumption that two-body collisional relaxation dominates the dynamics and that the cluster experiences a tidal field.
However, \citet{Williams2012} showed that for most of the regions of a massive star cluster, except for the central regions, collisionless relaxation should dominate over collisional relaxation. Consequently, they argue that the density profile of a GC should be better represented by DARKexp profiles \citep[][]{Williams_2010}, which are derived as the maximum entropy states of collisionless self-gravitating systems \citep[][]{Hjorth_2010}, than King profiles.

In terms of how the structure evolves, Hénon, in his thesis \citep{Henon_1961} and a subsequent article \citep{Henon_1965}\footnote{See \citet{Henon_2011} and \citet{Henon2_2011} for the corresponding English translations by Florent Renaud.}, studied the structure and evolution of a simplified star cluster in a tidal field and in isolation, respectively. He found that the final state of a tidally limited cluster is a homologous evolution\footnote{However, this is only true until a central source starts to provide energy to the system \citep{Henon_1975}}, i.e., evolution with a constant structure up to a scaling, with a continuous escape of high-energy stars resulting in a linear mass loss with time and a corresponding contraction due to the action of the tidal field. On the other hand, an isolated cluster evolves with a constant mass while expanding in size. \citet{Gieles2011} unified these two regimes into a unified set of equations that describe the evolution of a star cluster's tidal and half-mass radii, $\rhm$ during its balanced evolution. ``Balanced evolution" refers to the times when the energy production in the core balances the energy requirement to sustain the collapsing cluster.  

Several authors have focused on the structure of dense star clusters during its post core-collapse evolution. \citet{Cohn_1980} studied one-component star cluster models using numerical integration of the Fokker-Planck equation. They found that during late core-collapse evolution, the density profile becomes self similar with a power-law exponent of $\approx-2.2$, as argued by \citet{Lynden-Bell_1980}.
\citet{Giersz_1996} also found that the density profile after core-collapse evolves self-similarly. Their cluster models comprised stars with a power-law mass spectrum. \citet{Baumgardt_2002} also studied isolated star clusters, but with $N$-body models comprising equal-mass stars. They compared the evolution of models started with a Plummer  \citep{Plummer_1911} and a King profile with $\wk=3$. Interestingly, they found that the density profile during the expansion phase \emph{after} core-collapse is independent of the initial profile as well as the particle number. Similarly, \citet{Kupper_2008} showed that $\rhm$ after core collapse settles into a fixed value for cluster models with different initial $\rhm$. These models also comprised equal-mass stars without stellar evolution.

Traditionally, it was believed that any BHs that are retained in a GC after formation get dynamically decoupled and are quickly ejected \citep[e.g.,][]{Spitzer_1969,Kulkarni_1993}. However, this understanding has been updated in the past few years and it is now believed that not only can GCs retain a significant population of BHs to the present-day, they play a significant role in the structural evolution of GCs \citep[e.g.,][]{Morscher_2013,Morscher2015}. In fact, the observed non-core-collapsed GCs are believed to be supported against gravitational collapse by the energy production from BH binaries in their cores, a process called ``BH burning" \citep[e.g.,][]{Kremer_2020}. To that end, there have been efforts to understand the structural evolution of massive clusters during this ``pre" core-collapse phase as well.
\citet{Trenti2010} studied realistic star cluster $N$-body models via their observed core-to-half-light radius ratio. They found that this ratio, after a few relaxation times, is independent of the initial density profile and binary fraction. However, this equilibrium value does depend on the assumed fraction of retained dark remnants as they determine the amount of heating provided to the cluster model. \citet{Lamers2010} also notice that tidally under-filling cluster models, initialized with King profiles of differing $\wk$ tend to evolve to a $\wk=7$ profile. More recently, \citet{Sharma_2025} analyzed the density-profile evolution of massive star cluster models and found that models initialized with EFF \citep{Elson_1987} profiles naturally evolve to a structure best described as a King profile. Interestingly, they found that the final $\wk$ is independent of the initial profile.

In this paper, we also aim to study the structural evolution\footnote{Note that the ``structure" of a star cluster constitutes the positions as well as the velocities of its members, as the position alone can not fix the orbit of an object in the cluster potential.} of collisional dense star clusters during the pre-core-collapse phase. This is important to understand what, if anything, could be inferred about the initial properties of, e.g., GCs by observing their present-day surface brightness profiles. In \autoref{sec:methods}, we describe our model set and assumptions. In \autoref{sec:results}, we present our results. We find that our cluster models exhibit an ``equilibrium" density profile to which they converge, irrespective of their initial structure, after $\sim\mathcal{O}(\gyr)$. In \autoref{sec:exp}, we present the results of a numerical experiment we performed with a set of purely dynamical models to check the ``stability" of this equilibrium profile. In \autoref{sec:discuss}, we try to make sense of our results from a theoretical perspective. Finally, in \autoref{sec:summ}, we conclude our study.

\section{Methods}
\label{sec:methods}

To study the structural evolution of dense star clusters, we simulate several model star clusters using the Hénon-type \citep{Henon_1971a,Henon_1971b} Monte Carlo, star cluster evolution code, $\cmc$ \citep{Joshi_2000,Joshi_2001,Chatterjee2010,Goswami_2012,Morscher_2013,Pattabiraman_2013,Rodriguez2022}. The particles in $\cmc$ represent individual stars, which evolve under all the important physical processes believed to be necessary to accurately model a dense star cluster's evolution. The stellar and binary evolution is handled by the population synthesis code, $\cosmic$ \citep{Breivik_2020}, which in turn utilizes the $\texttt{BSE}$ package \citep{Hurley2000,Hurley2002}. Two-body relaxation, which is a cumulative effect of several weak encounters over many orbits, is modeled using a single ``super-encounter" per timestep for all the pairs of stars/binaries. Strong 3/4-body encounters are modeled using the small-$N$ gravitational dynamics package, $\texttt{fewbody}$ \citep{Fregeau2007}. In the parameter space where the assumptions of the Monte Carlo method are expected to be valid, $\cmc$ has shown good agreement with direct $N$-body codes \citep[e.g.,][]{Morscher2015,Rodriguez_2016,Khurana_2025}.

\begin{table*}
    \centering
    \begin{tabular}{cccccccc}
        Model (set) & $N$ & $M$ ($\msun$) & $\wk$ & $\rv$ ($\pc$) & $\fb$ ($\%$) & $\rtid$ ($\pc$) & $\trhm$ ($\gyr$)\\ \hline \hline
        Fiducial & $8\times10^5$ & $\approx4.8\times10^5$ & $8$ & $1.0$ & $5$ & $\approx130$ & $\approx0.21$\\ \hline
        \multirow{4}{*}{King} & .. & .. & $0.1$ & .. & .. & .. & ..\\
         & .. & .. & $4$ & .. & .. & .. & ..\\
         & .. & .. & $12$ & .. & .. & .. & $\approx0.25$\\
         & .. & .. & $18$ & .. & .. & .. & $\approx0.25$\\ \hline
        \multirow{3}{*}{EFF} & .. & .. & $\gamma=2.5$ & .. & .. & .. & $\approx0.17$\\ 
         & .. & .. & $\gamma=4$ & .. & .. & .. & $\approx0.18$\\
          & .. & .. & $\gamma=9$ & .. & .. & .. & $\approx0.20$\\ \hline
        \multirow{7}{*}{Vary $\fb$} & .. & $\approx4.7\times10^5$ & .. & .. & $0$ & .. & ..\\
         & .. & $\approx5.4\times10^5$ & .. & .. & $25$ & .. & ..\\
         & .. & $\approx6.0\times10^5$ & .. & .. & $50$ & .. & $\approx0.19$\\
         & .. & $\approx6.6\times10^5$ & .. & .. & $70$ & .. & $\approx0.18$\\
         & .. & $\approx6.8\times10^5$ & .. & .. & $80$ & .. & $\approx0.18$\\
         & .. & $\approx7.4\times10^5$ & .. & .. & $99$ & .. & $\approx0.17$\\
         & .. & $\approx7.4\times10^5$ & $4$ & .. & $99$ & .. & $\approx0.16$\\
         & .. & $\approx7.4\times10^5$ & $12$ & .. & $99$ & .. & ..\\ \hline
        \multirow{3}{*}{Compact} & .. & .. & $0.1$ & $0.5$ & .. & .. & $\approx0.07$\\
         & .. & .. & .. & $0.5$ & .. & .. & $\approx0.08$\\
         & .. & .. & $18$ & $0.5$ & .. & .. & $\approx0.09$\\ 
    \end{tabular}
    \caption{The initial properties of the models we have simulated in this study. The columns are: (1) the name of the model (set), (2) the number of stars/binaries, $N$, (3) the mass of the cluster, $M$, (4) the King $\wk$ parameter (for the EFF model set, we provide the $\gamma$ instead), (5) the virial radius, $\rv$, (6) the binary fraction, $\fb$, (7) the tidal radius, $\rtid$, and (8) the half-mass relaxation time, $\trhm$. The first row defines the fiducial model, and the rows 2-5 and 9-16 define the model sets with varying $\wk$ and $\fb$, respectively. The fiducial model is a part of both these model sets. The rows 6-8 define the set of  models with EFF profiles. The rows 17-19 define the set of compact models with $\rv=0.5\,\pc$. ``.." denotes that the value is the same as the fiducial model.}
    \label{tab:sims}
\end{table*}

The star cluster models that we have simulated for this study can be categorized into several model sets. These sets are created by changing a particular cluster property with respect to a constant ``fiducial" model, unless otherwise noted. The fiducial model has an initial number of objects (singles/binaries), $N=8\times10^5$ and a binary fraction, $\fb=5\%$. The stellar masses are drawn from a \citet{Kroupa2001} mass function ranging from $0.08\,\msun$ to $150\,\msun$. This results in an initial cluster mass, $M\approx4.8\times10^5\,\msun$. For the binaries, orbital periods are drawn from a log-uniform distribution ranging from the Roche-lobe overflow condition to the local hard-soft boundary \citep[e.g.,][]{Abt_1983}. The eccentricities are distributed thermally, i.e. $dN/de = 2e$ \citep[e.g.,][]{Jeans_1919}. The mass ratios are sampled uniformly between 0 and 1. The metallicity of the stars is fixed at $Z=0.001$ while the solar metallicity is assumed to be $Z_\odot=0.017$. The objects' positions and velocities follow a King profile with $\wk=8$ and a virial radius, $\rv=1\,\pc$. Finally, the cluster is assumed to be orbiting the Milky Way in a circular orbit at a radius of $10\,\kpc$. This corresponds to a tidal radius, $\rtid\approx130\,\pc$.

Now, the first set of models has varying initial density profiles, characterized by $\wk$, while keeping everything else the same as the fiducial model. This set includes 5 models, with $\wk=0.1,\ 4,\ 8\ (\rm{fiducial}), 12,\ \rm{and}\ 18$. We found $\wk=0.1$ and $18$ to be near extreme choices for a King profile, with central density increasing monotonically with $\wk$. In addition, we also include three models with an EFF profile with $\gamma=2.5$, $4$ (i.e., Plummer) and $9$.

Another set of models has a varying $\fb$. These include $\fb=0,\ 5\ (\rm{fiducial}),\ 25,\ 50,\ 70,\ 80,\ \rm{and}\ 99\%$. As the number of objects are the same as the fiducial model, i.e., $N=8\times10^5$, the total cluster mass increases with $\fb$. This is because increasing $\fb$ keeping the number of objects constant leads to an increase in the total number of individual stars, $N_{\rm{star}}=N(1+\fb)$ in the model. So, where $N_{\rm{star}}=8\times10^5$ for the $\fb=0\%$ model, it is $1.2\times10^6$ for the $\fb=50\%$ model. The corresponding masses in $\msun$ for the aforementioned models are $\approx4.7\times10^5$, $\approx4.8\times10^5$ (fiducial), $\approx5.4\times10^5$, $\approx6.0\times10^5$, $\approx6.6\times10^5$, $\approx6.8\times10^5$, and $\approx7.4\times10^5$. For $\fb=99\%$, we simulate two additional models with $\wk=4$ and $12$.

In addition, we also simulated a set of relatively compact models with a smaller $\rv$ of $0.5\,\pc$. Since the binding energy,
\begin{equation}
\label{eq:E}
    |E|=GM^2/4\rv
\end{equation}
\citep[e.g.,][]{HeggieHut_2003}, where $M$ is the cluster mass and $G$ is the gravitational constant, these models have twice the binding energy of the fiducial model. In this set, we simulated three models with $\wk=0.1,\ 8,$ and $18$.

\autoref{tab:sims} summarizes the initial properties of all the models we have simulated for this study. In the last column, we also provide the initial half-mass relaxation time, $\trhm$ for these models.

\section{Results}
\label{sec:results}

\subsection{Varying the initial profile}
\label{subs:w0}

\begin{figure}
    \centering
    \includegraphics[width=1.0\linewidth]{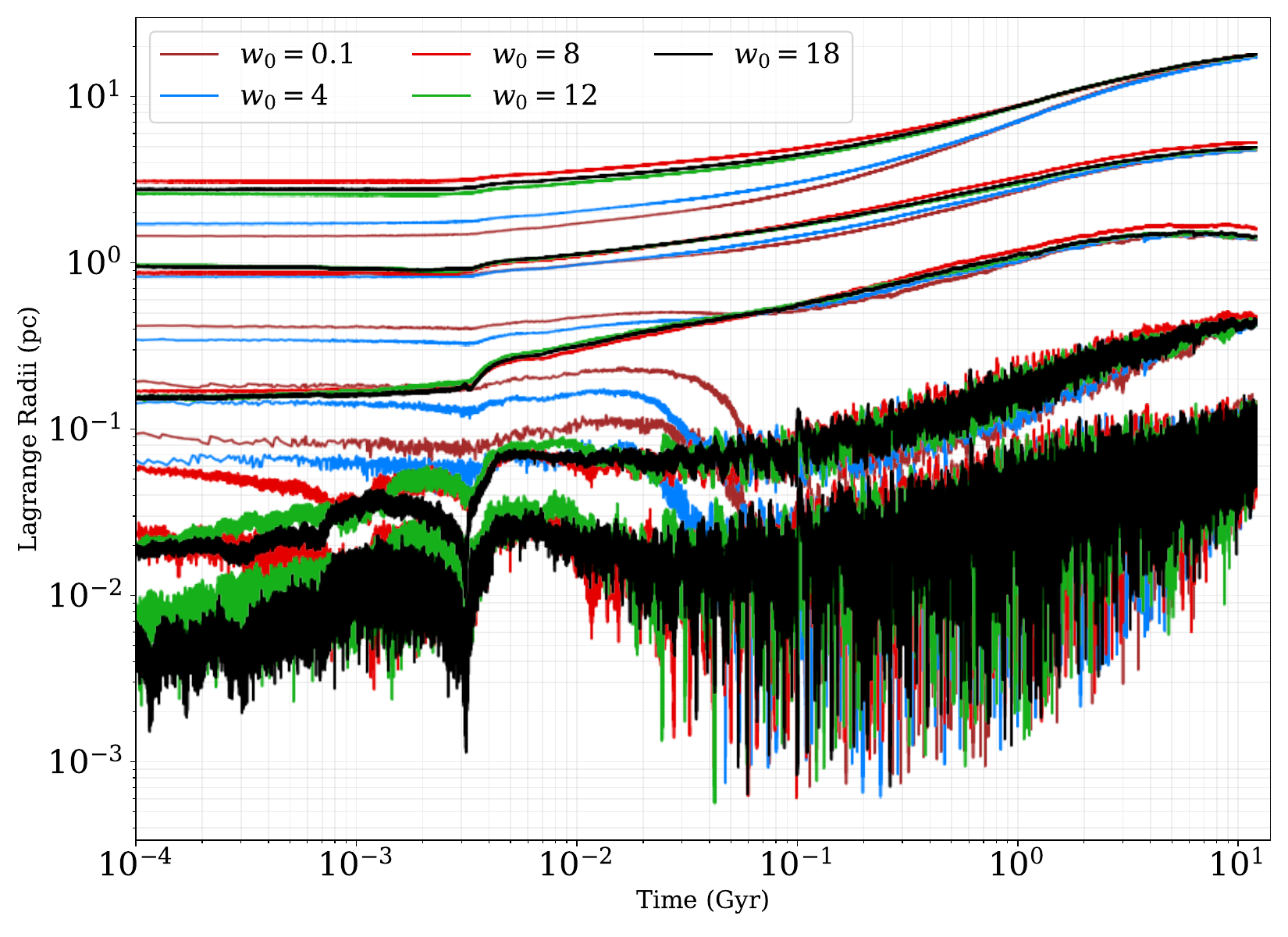}
    \caption{The 0.1, 1, 10, 50, and 90\% Lagrange radii (from bottom to top) of the King model set as a function of time. Colors of the curves represent the initial value of $\wk$.}
    \label{fig:kingw0}
\end{figure}

\begin{figure}
    \centering
    \includegraphics[width=1.0\linewidth]{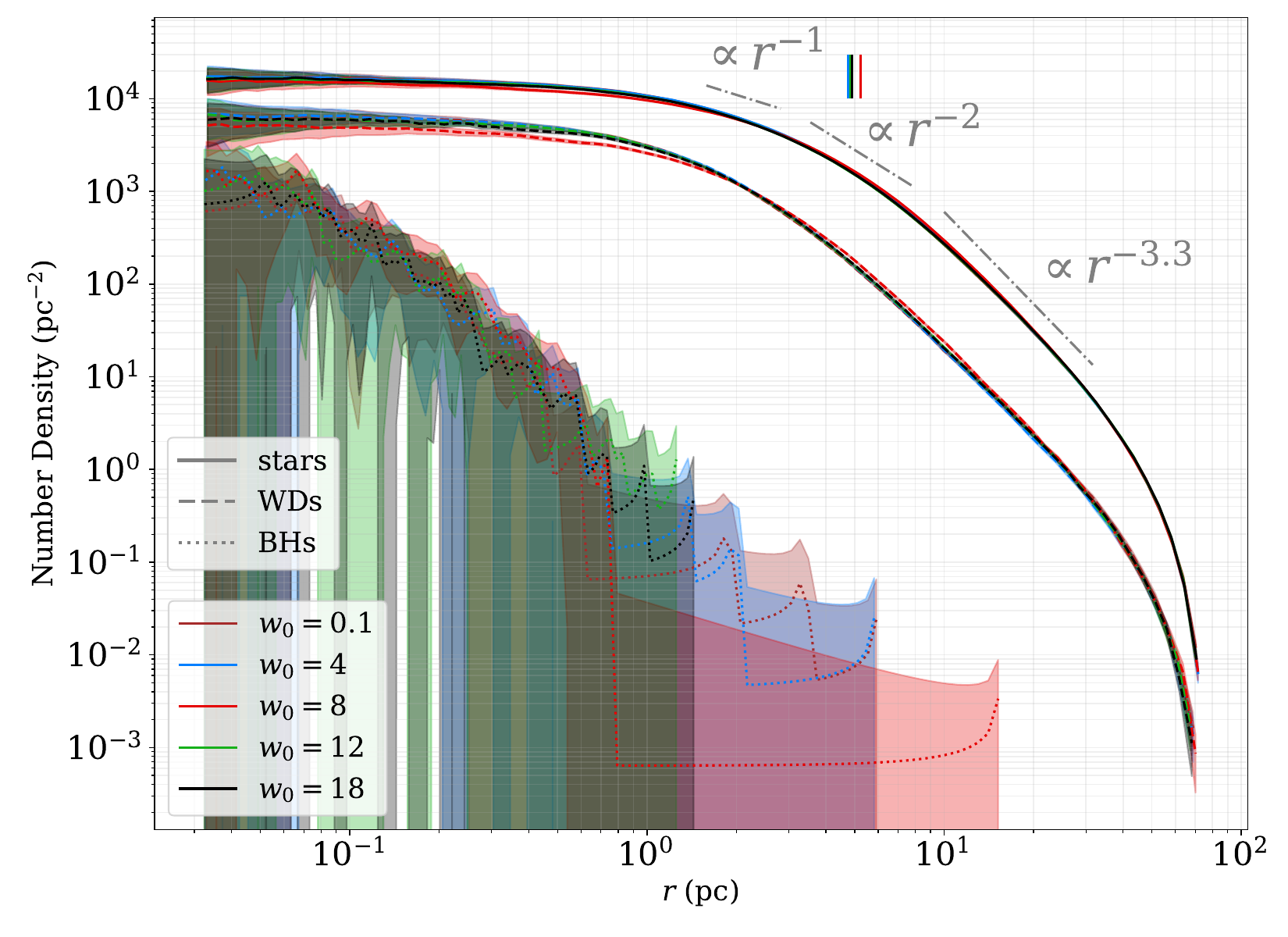}
    \caption{The projected number density profiles of the King model set at $t=12\,\gyr$, created using \texttt{cmctoolkit}. The solid, dashed, and dotted curves represent non-compact-object stars, white dwarfs, and BHs, respectively. The shaded regions denote the Poisson uncertainty. The gray dashed-dotted line segments show sample power-law curves $\propto r^{-1},\ r^{-2}$, and $r^{-3.3}$. The colored vertical line segments denote the respective current $\rhm$s for the models.}
    \label{fig:numden}
\end{figure}

\begin{figure}
    \centering
    \includegraphics[width=1.0\linewidth]{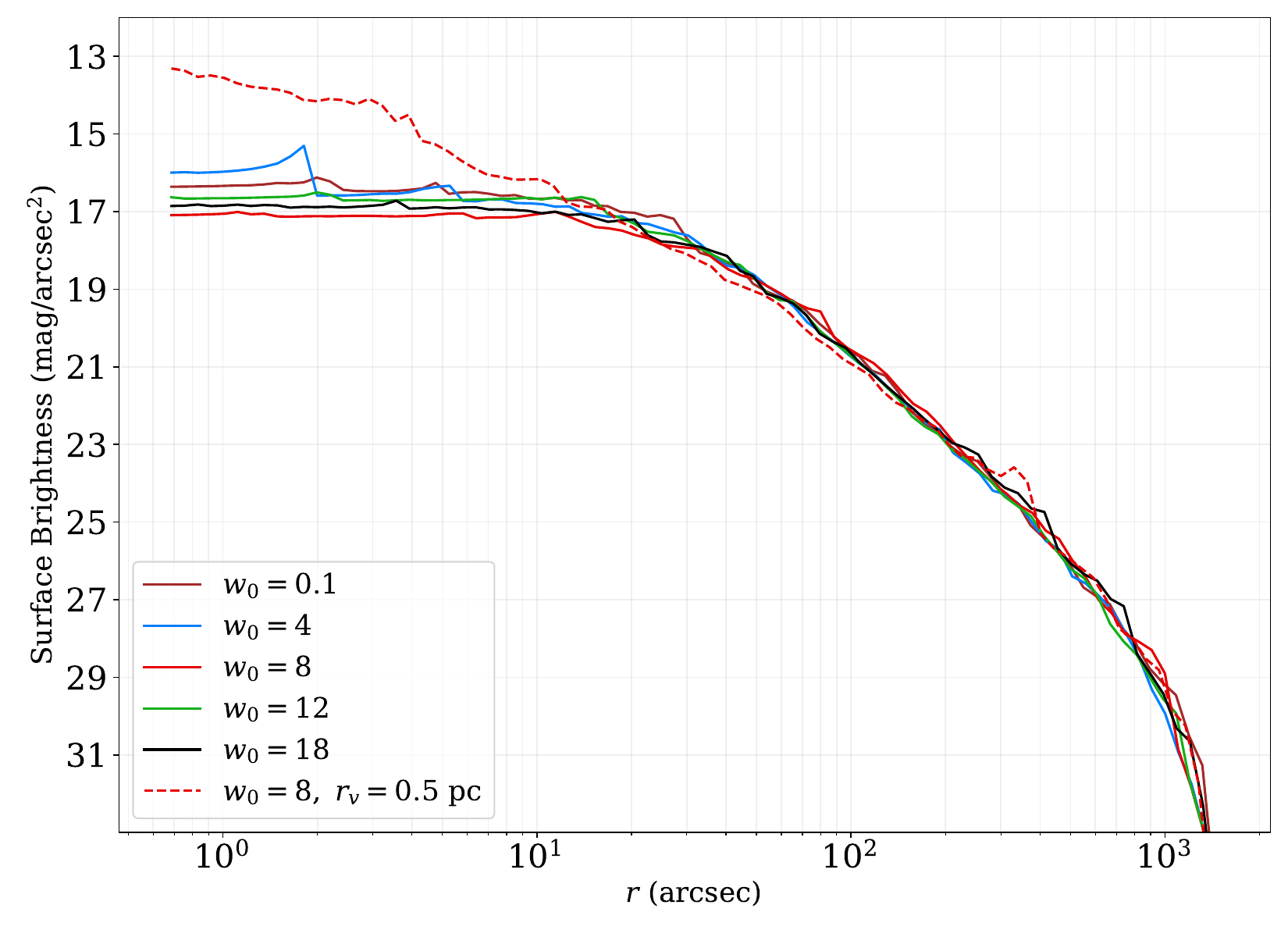}
    \caption{The mock-observed $V$-band surface brightness profiles of the King model set at $t=12\,\gyr$, created using \texttt{cmctoolkit} (solid curves). The dashed red curve represents the model with $\wk=8$, but $\rv=0.5\,\pc$ (\autoref{subs:rv}). The model clusters are assumed to be at a distance of $10\,\kpc$ from the sun.}
    \label{fig:magV}
\end{figure}

In \autoref{fig:kingw0}, we plot the evolution of the structure of the model clusters with different initial density profiles by means of various Lagrange radii, $\rl$s ($x\%\ \rl\equiv$ radius of the 3-D sphere enclosing $x\%$ of the total mass). During the initial stages of evolution ($\lesssim3\,\myr$), the outskirts of all models remain largely unchanged. However, the central regions evolve. The nature of this evolution depends on the initial $\wk$. The less dense models, i.e., $\wk\lesssim 8$ contract due to two-body relaxation and mass segregation, while the denser models expand \citep[e.g.,][]{Lamers2010}. At $\sim3\,\myr$, when the most massive stars start exploding as supernovae (SNe), all the models expand due to stellar evolution driven mass-loss. After the mass-loss rate due to stellar evolution becomes small \citep[$\sim35\,\myr$, e.g.,][]{Khurana_2025}, all models go into the quasi-stable state of BH burning \citep[e.g.,][]{Kremer_2020,Weatherford2021}. During this stage of the evolution, we observe that all the models, with vastly different initial density profiles, tend to converge to very similar density profiles. We call this final profile an ``equilibrium" profile. The inner $\rl$s quickly ($\lesssim 100\,\myr$) become very similar, while the $90\%$ $\rl$s take a few $\gyr$ to follow. In any case, by the time the simulations end, all the models show very similar density profiles.

In \autoref{fig:numden}, we show the projected number density profiles of these models, divided into non-compact-object stars, white dwarfs, and BHs, at the end of the simulation, i.e., at $t=12\,\gyr$, created using \texttt{cmctoolkit} \citep[][]{Rui2021,rui_2021}. All profiles for models started with different King $\wk$ are almost identical. The radial dependence of the intermediate regions outside the flat core is close to a power-law $\propto r^{-3.3}$. However, if we focus on the region around $\rhm$s, the radial dependence is close to $\propto r^{-\{1<\alpha<2\}}$.

In \autoref{fig:magV}, we also show, as solid lines, the mock-observed $V$-band \citep{Rodrigo_2020} surface brightness profiles for the models with different King profiles at the last snapshot, i.e., at $t=12\,\gyr$, also created using \texttt{cmctoolkit}. We include light from all stars except for compact objects, i.e., white dwarfs, neutron stars, and BHs. Furthermore, to calculate the angular scale of the $x-$axis, the model clusters are assumed to be at a distance of $10\,\kpc$. This figure shows the similarity of the structural profile of our models from an observational perspective. Note that the Lagrange radii in \autoref{fig:kingw0} take into account the mass of all stars, including the compact objects. Thus, \autoref{fig:numden} and \autoref{fig:magV} present complementary evidence for the similarity of model clusters' structure at $t=12\,\gyr$.

\begin{figure}
    \centering
    \includegraphics[width=1.0\linewidth]{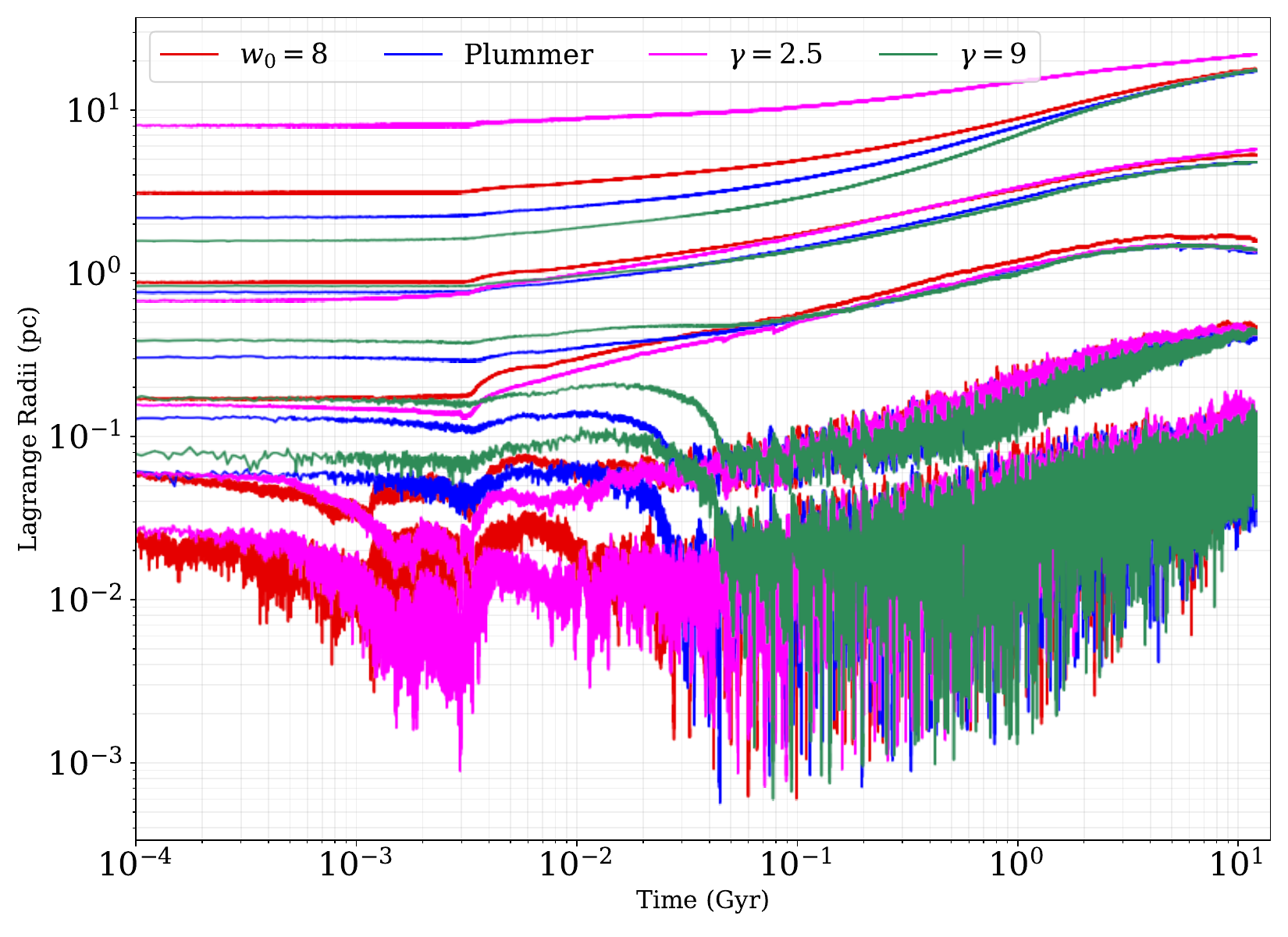}
    \caption{The same as \autoref{fig:kingw0}, except here we compare the fiducial model to models with EFF (with $\gamma=2.5$ and $9$) and Plummer initial density profiles (or equivalently, an EFF profile with $\gamma=4$).}
    \label{fig:eff}
\end{figure}

Similarly, in \autoref{fig:eff}, we compare how models with EFF and Plummer profiles evolve compared to the fiducial model. Again, we observe that irrespective of the initial density profile, the cluster models tend to the equilibrium profile during BH burning.

\subsection{Varying the initial binary fraction}
\label{subs:fb}

\begin{figure}
    \centering
    \includegraphics[width=1\linewidth]{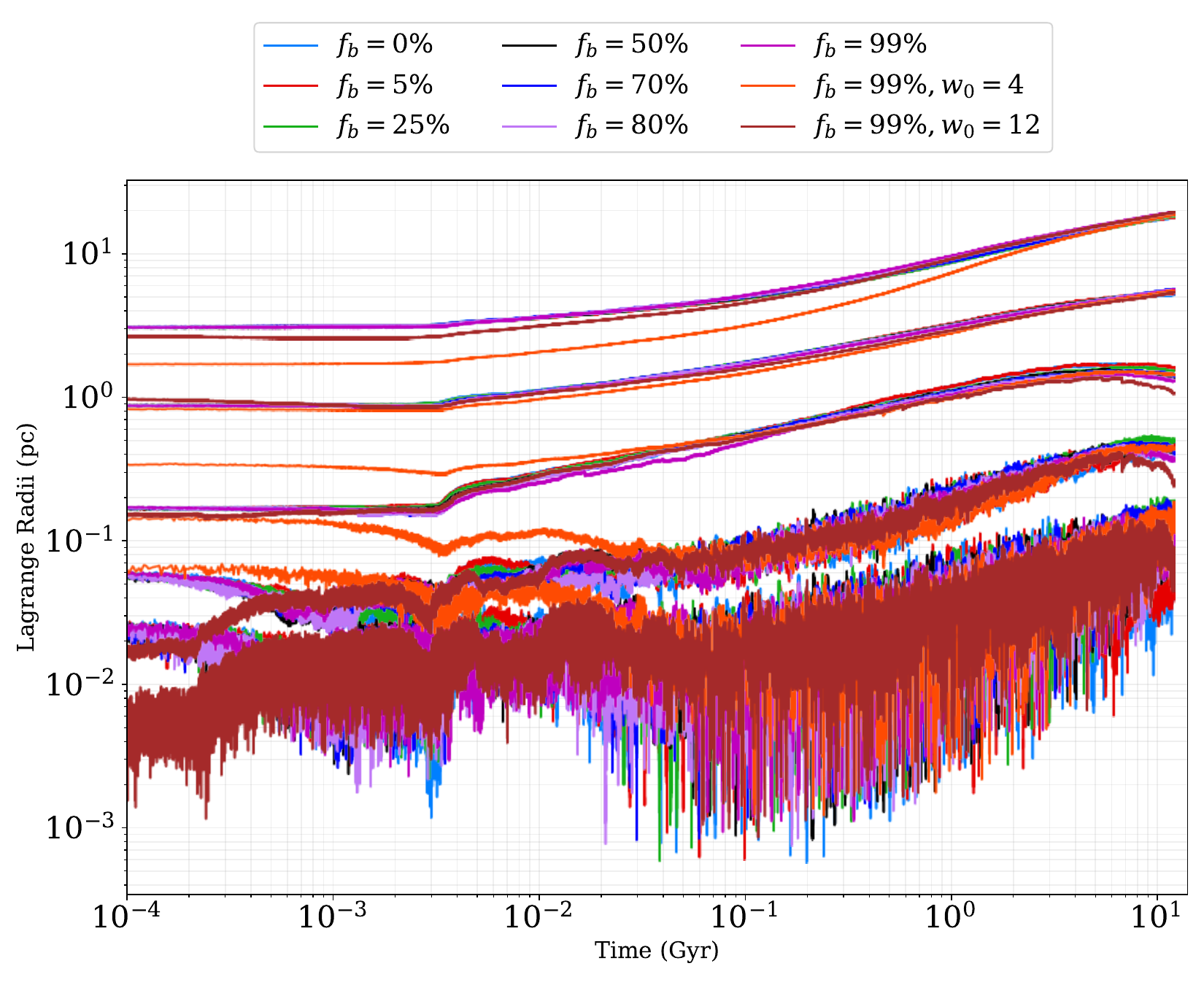}
    \caption{The same as \autoref{fig:kingw0}, except here we compare models with the same initial density profile, i.e., $\wk=8$, but different initial $\fb$. For $\fb=99\%$, we show two additional models with $\wk=4$ and $12$.}
    \label{fig:fb}
\end{figure}

In \autoref{fig:fb}, we compare the structural evolution of the models with different initial $\fb$. Notice that the $\fb=5\%$ model (red) in this figure is the same as the $\wk=8$ model in \autoref{fig:kingw0} and \autoref{fig:eff}, i.e., our fiducial model. We observe that all of these models also attain the equilibrium profile. Also, since seven of the models shown in this figure have the same initial density profile, they overlap each other to such an extent that the $\fb=99\%$ model nearly hides the lower-$\fb$ models. Thus, the convergence to an equilibrium profile during BH burning does not depend even on the initial $\fb$. 
Note that unlike the models in \autoref{fig:kingw0}, these models do not have the same initial mass or binding energy (see \autoref{sec:methods}). 

Also note that for $\fb=99\%$, \autoref{fig:fb} shows three models, with $\wk=8,\ 4$ and $12$. Even their convergence to the equilibrium profile is further evidence that the equilibrium profile is independent of the initial $\fb$ and $\wk$.

\subsection{A set of compact models}
\label{subs:rv}

\begin{figure}
    \centering
    \includegraphics[width=1\linewidth]{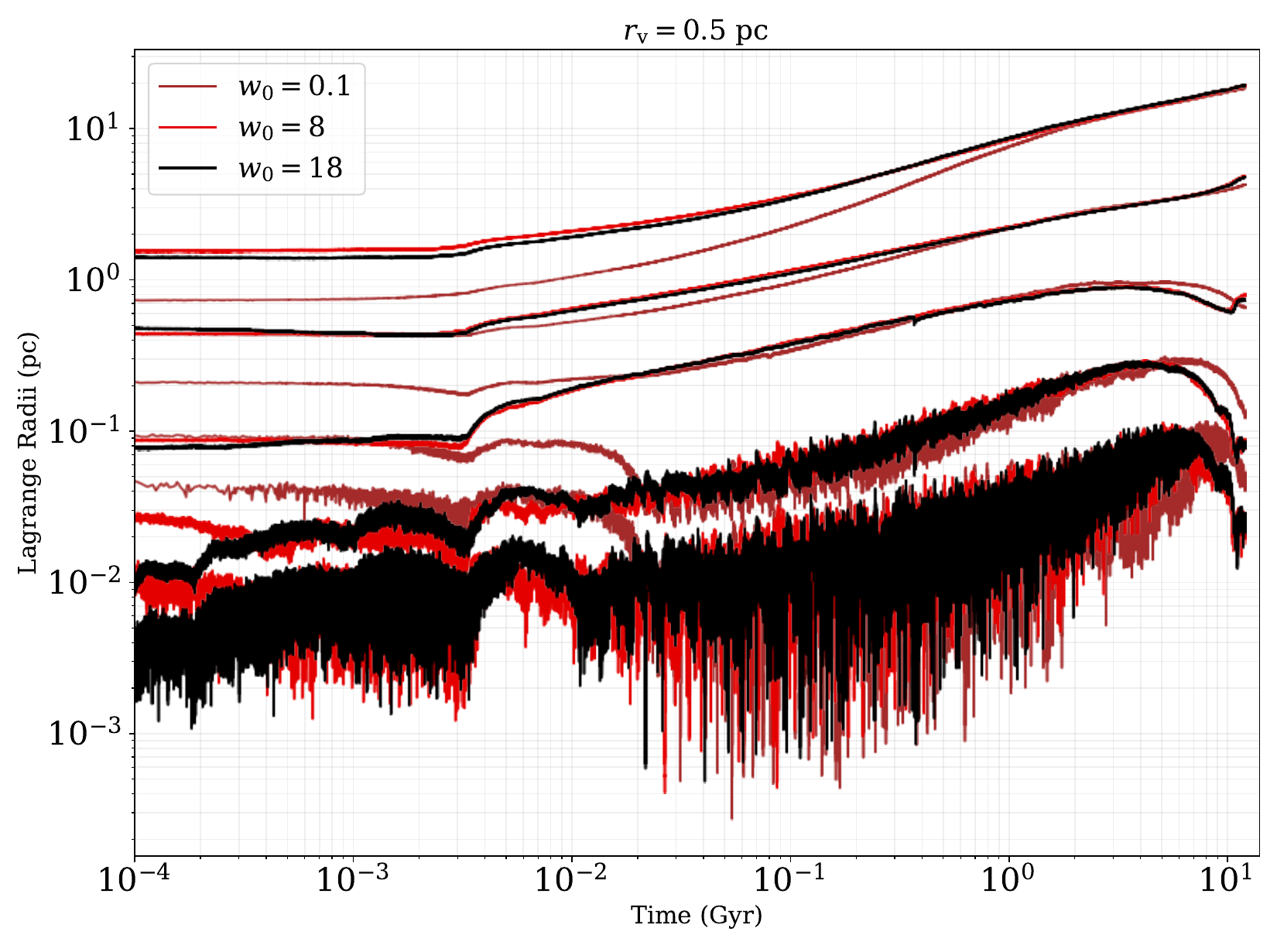}
    \caption{Same as \autoref{fig:kingw0}, except here we compare relatively compact cluster models with $\rv=0.5\,\pc$.}
    \label{fig:rv05}
\end{figure}

\begin{figure}
    \centering
    \includegraphics[width=1\linewidth]{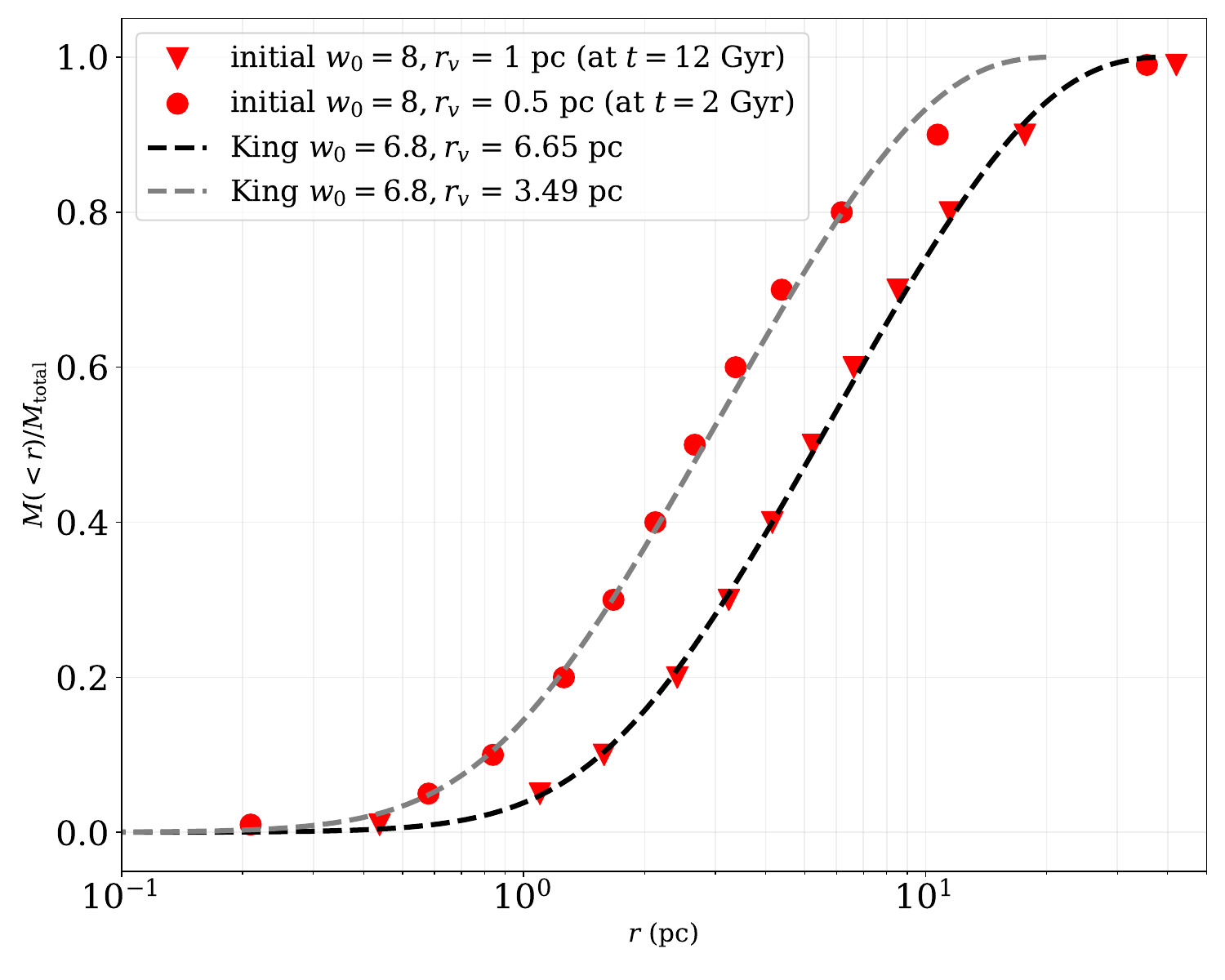}
    \caption{The equilibrium profiles for the model sets with $\rv=1$ and $0.5\,\pc$, shown as cumulative mass distributions. The red markers represent the data extracted from $\cmc$ for the models with initial $\wk=8$ from both the model sets. For the $\rv=1\,\pc$ model, the data is from the $t=12\,\gyr$ snapshot, while for the $\rv=0.5\,\pc$ model, it is from the $t=2\,\gyr$ snapshot. Both the model sets exhibit the equilibrium profiles during these respective times (see \autoref{fig:kingw0} and \autoref{fig:rv05}). The black and gray curves represent King profiles with $\wk=6.8$, but with an $\rv=6.65$ and $3.49\,\pc$, respectively.}
    \label{fig:fprof}
\end{figure}

To further check the limits of the tendency of models to converge to an equilibrium profile, we simulated a set of models with $\rv=0.5\,\pc$ and $\wk=0.1,\ 8,$ and $18$. All other properties are similar to the fiducial model. In \autoref{fig:rv05}, we show the evolution of these models. Again, these models converge to an equilibrium profile during BH burning. 

However, we observe that the profiles in \autoref{fig:rv05} start to diverge from each other during the last few $\gyr$. This deep collapse, observed as a drop in inner $\rl$s, occurs when the cluster model ejects almost all of its BHs, as BH binaries were the dominant source of energy keeping the cluster from collapsing before. After BHs are ejected, stellar binaries, along with possibly some white dwarfs, replace BH binaries as the energy source at the center. Being orders of magnitude less energetic, stellar binaries need much higher central density to sustain the cluster model \citep[e.g.,][]{Mackey2008}. This is believed to be the state of GCs observationally classified as core-collapsed \citep[e.g.,][]{Chatterjee2013}. Furthermore, after BHs are ejected, luminous stars can populate the central regions of the cluster models. The resulting difference in the mock-observed surface brightness profile during this phase compared to the BH-burning phase can be seen in \autoref{fig:magV}. The model with $\wk=8$ from this set at $t=12\,\gyr$ is represented by the dashed red line. The characteristic cusp close to the cluster center associated with core-collapse is clearly discernible.

In \autoref{fig:fprof}, we focus on the equilibrium profiles exhibited by the model sets with $\rv=1$ (at $t=12\,\gyr$) and $0.5\,\pc$ (at $t=2\,\gyr$). To find a theoretical profile that matches the equilibrium profiles, we first calculate the $\rv$ of the models during the aforementioned times using \autoref{eq:E}. Then, we sweep $\wk$ to visually find a matching profile. Interestingly, we find that both sets of models are well represented by a $\wk=6.8$ King profile, just scaled by their respective $\rv$s. \citet{Lamers2010} also note that tidally under-filling models initialized with a higher or lower $\wk$ readjusts its structure towards a $\wk=7$ profile upon expansion in its Roche lobe. However, the outskirts of the $\rv=0.5\,\pc$ model are more extended compared to the theoretical fit. This might be because the initial $\rtid$ is the same for both models (\autoref{tab:sims}), and the relatively compact models have more room to expand compared to the $\rv=1\,\pc$ models.

\section{Profile restoration experiment}
\label{sec:exp}

We have seen in the previous section that the dynamical evolution of collisional star cluster models seems to exhibit an equilibrium density profile, given their initial mass and binding energy. Now, we aim to check the ``stability" of this equilibrium. In other words, we perform a numerical experiment in which we artificially perturb the density profile, while keeping the binding energy constant, of a cluster model during the equilibrium phase, to check if it restores. 

\begin{figure}
    \centering
    \includegraphics[width=1\linewidth]{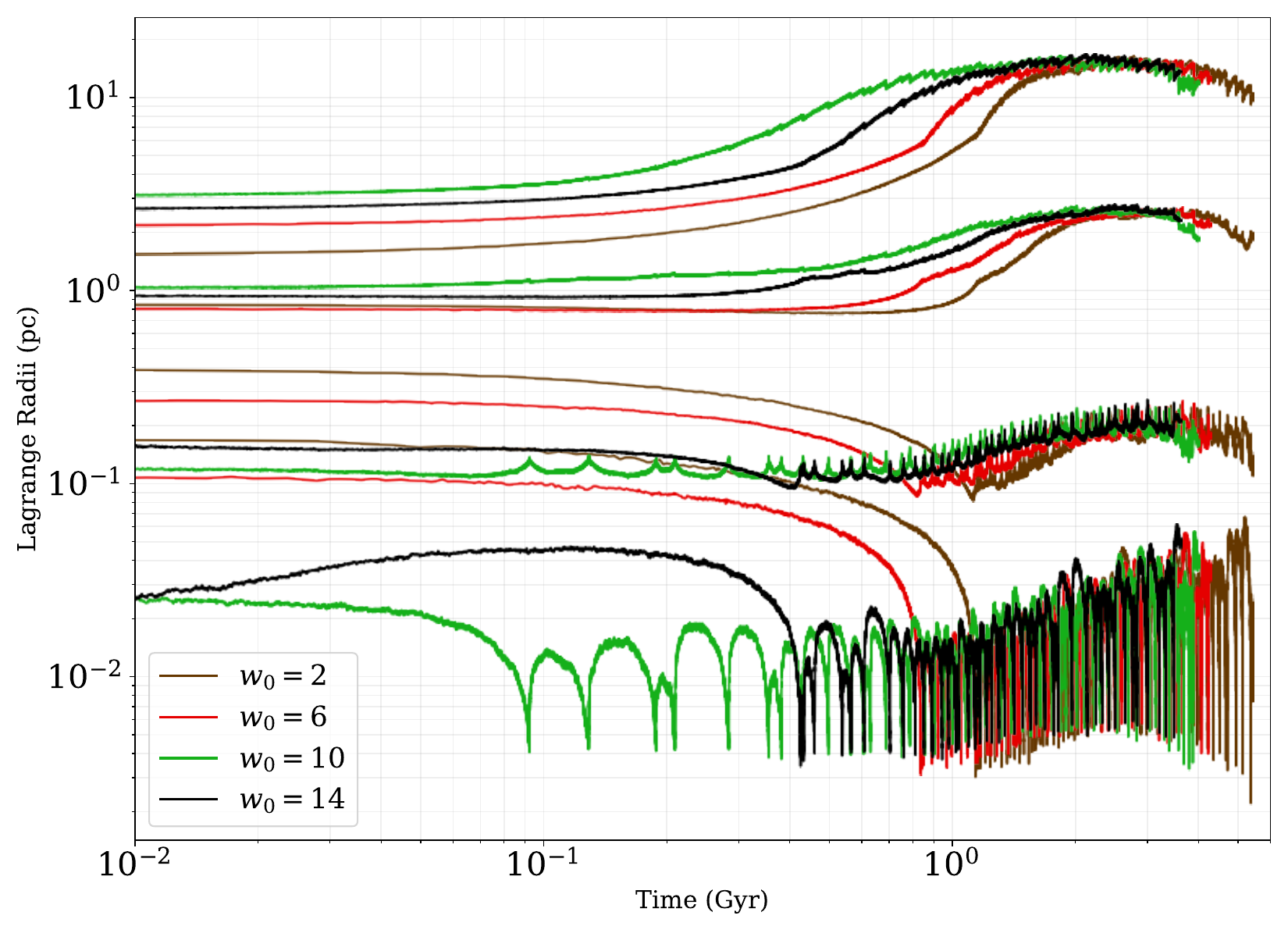}
    \caption{The time evolution of the 1, 10, 50, and 90\% $\rl$s (from bottom to top) for the set of purely dynamical models.}
    \label{fig:dyn1}
\end{figure}

\begin{figure}
    \centering
    \includegraphics[width=1\linewidth]{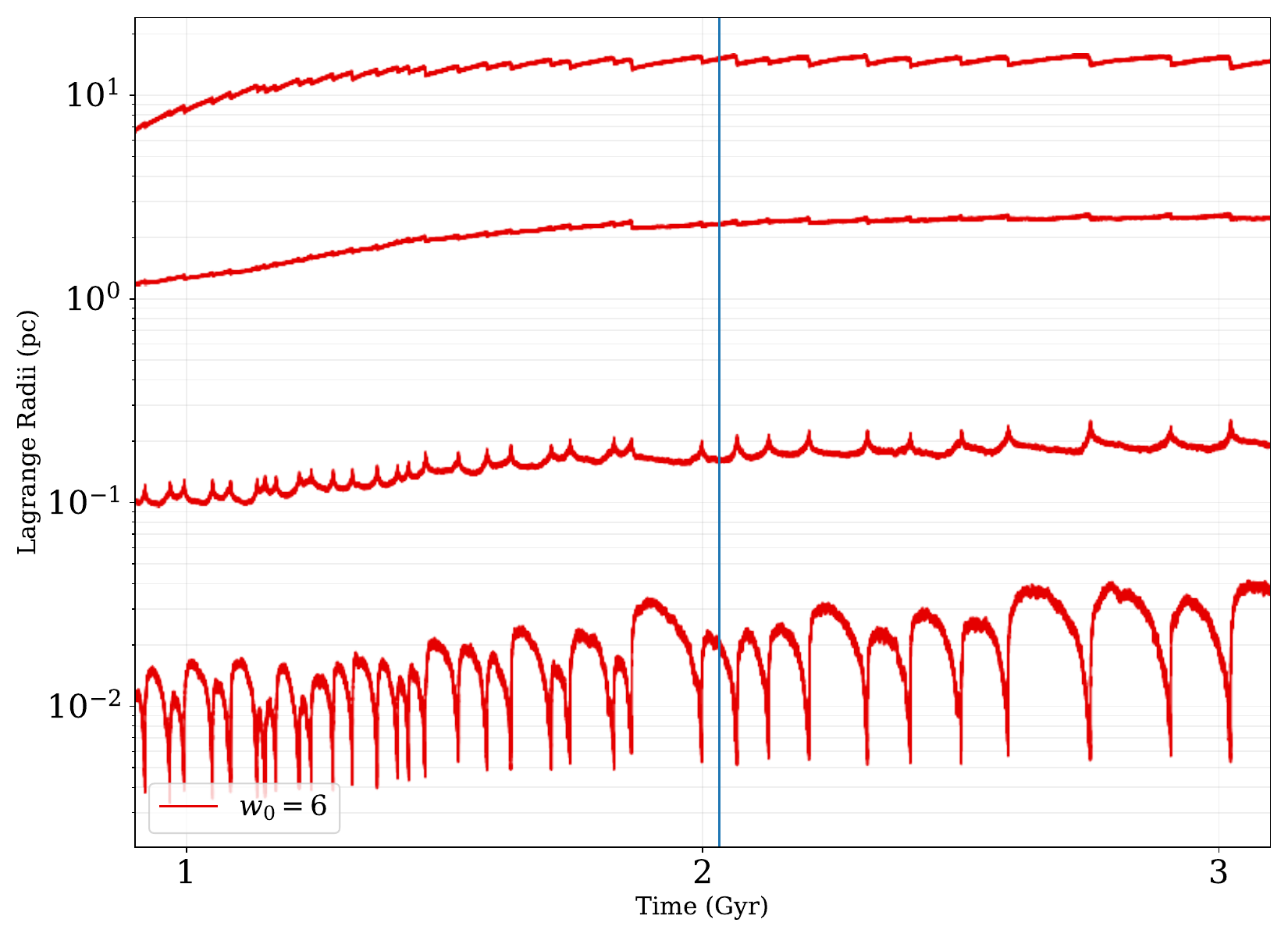}
    \includegraphics[width=1\linewidth]{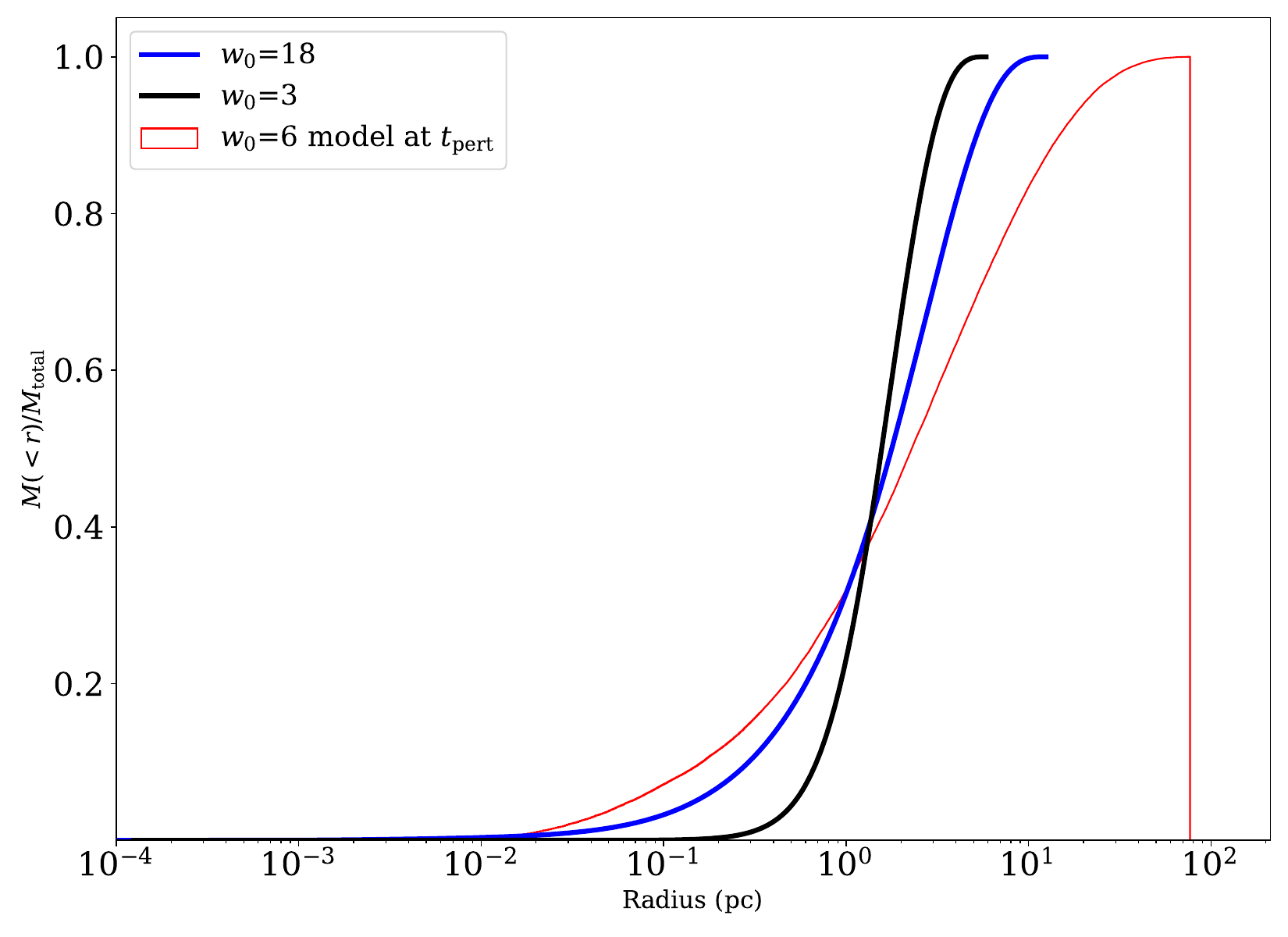}
    \caption{\emph{Top}: An $x$-axis zoom-in of \autoref{fig:dyn1} for the model with $\wk=6$. The vertical blue line represents the time-coordinate, ($\tpert\sim2\,\gyr$) at which we choose to perturb the profile to create two additional models. \emph{Bottom}: The density profiles of the original model at $\tpert$ and of the perturbed models. Red histogram represents the original profile corresponding to the vertical blue line in the top panel, while the black and blue curves represent the two perturbed profiles, characterized by $\wk=3\ \&\ 18$, respectively.}
    \label{fig:dynchoose}
\end{figure}

To isolate the effect of stellar dynamics, we perform this experiment in purely dynamical models without stellar evolution. For this purpose, we simulate a set of models with $10^5$ equal-mass point particles with no primordial binaries and various King profiles. However, dynamical formation of binaries via three-body interactions is allowed \citep[e.g.,][]{Morscher2015}. $\rv$ is set to $1.0\,\pc$ and $\rtid$ to $120\,\pc$. In \autoref{fig:dyn1}, we show the structural evolution of this set of models. Notice that these models dissolve within $\sim$ a few $\gyr$. \autoref{fig:dyn1} confirms that even purely dynamical models exhibit an equilibrium profile just as our realistic models. Again, the inner $\rl$s converge quickly, while the $90\%$ $\rl$ and the $\rhm$ converge after $\sim2\,\gyr$. All $\rl$s evolve similarly afterwards.

Now that we have established the existence of an equilibrium density profile in purely dynamical models as well, we choose a model snapshot at some time during the equilibrium phase and artificially perturb its profile. Then we let the new models with perturbed profiles evolve. The top panel in \autoref{fig:dynchoose} shows the state of the model with $\wk=6$ around the chosen time, $\tpert\sim2\,\gyr$, while the bottom panel shows the chosen density profile of this model in comparison to the artificially perturbed density profiles. We create two models with perturbed profiles, one with a higher ($\wk=18$), and another with a lower ($\wk=3$), central density. The two models with perturbed profiles preserve the total binding energy and mass of the original model at $\tpert$. We also make sure to preserve the cluster-centric distance ordering of all the objects (singles/binaries). This ensures that if the binaries are more segregated towards the cluster center in the original model, they remain so in the perturbed models as well. Now, we simulate the dynamical evolution of the perturbed models and check if they restore their structure to match the equilibrium profile as time passes.

\begin{figure}
    \centering
    \includegraphics[width=1\linewidth]{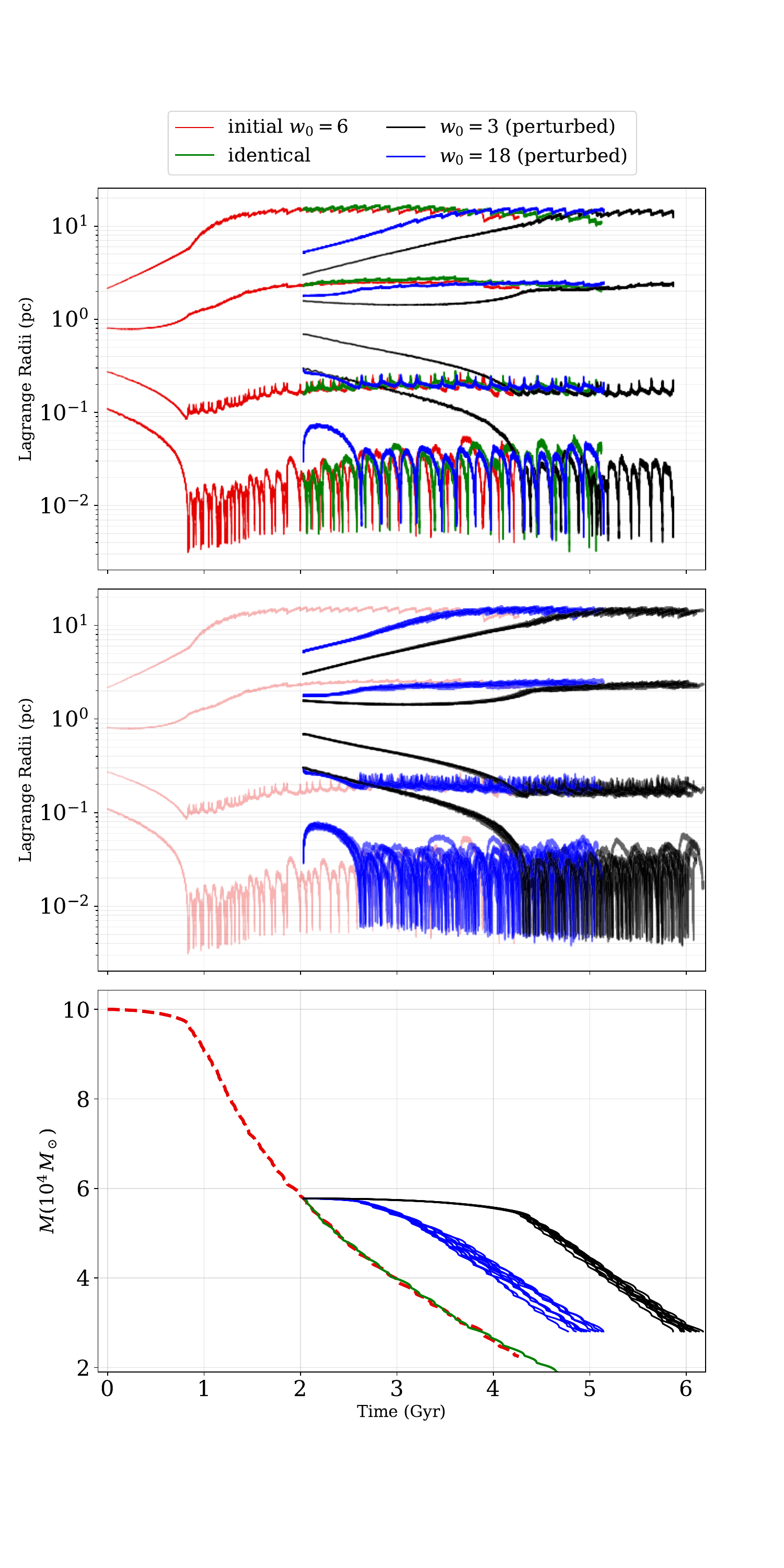}
    \caption{\emph{Top}: The 1, 10, 50, and 90\% $\rl$s (from bottom to top) of the two perturbed models ($\wk=3$ and $\wk=18$) as a function of time in comparison to the original model (initial $\wk=6$). \emph{Middle}: The same as the top panel, but here we include 10 different realizations of the two perturbed models to visualize the statistical fluctuations. \emph{Bottom}: The mass evolution of the same models. All the models are evolved until almost complete dissolution. In addition, in green, we show the evolution of a model that is restarted with identical member positions and velocities as the original model.}
    \label{fig:pert}
\end{figure}

In \autoref{fig:pert}, we show the evolution of the two perturbed models in comparison to the original model. In addition, we show the evolution of a model that is restarted with the member positions and velocities identical to the original model at $\tpert$, in green. Notice that the perturbed models begin at $\tpert$. We observe that the identical model evolves statistically similar to the original model, both in terms of the spatial structure as well as mass. This serves as a validation of our restart procedure for models using $\cmc$.
Thus, any differences in the evolution of the perturbed models could be safely attributed solely to the different initial profiles.

Returning our attention to the perturbed models in the top panel of \autoref{fig:pert}, we observe that they start their evolution at $\tpert$ with a significantly different structure than the original model. However, all the $\rl$s of both perturbed models almost immediately evolve towards the $\rl$s of the original model. In $\sim$ a few $\gyr$, both models settle into a very similar profile to the original model. As expected, the denser model ($\wk=18$) evolves quicker due to a smaller central relaxation time, while the less dense model ($\wk=3$) lags behind. To check the robustness of this result, we simulate 10 different realizations of both perturbed models by changing $\cmc$'s random-number-generator seed. The structural evolution of all of these realizations is plotted in the middle panel of \autoref{fig:pert}. It is evident that our result that the models with perturbed profiles should converge to the equilibrium profile is robust against statistical fluctuations.

A similar trend is also observed in mass evolution (bottom panel, \autoref{fig:pert}), where the rate of mass loss (the slope of mass-vs-time) of the perturbed models takes a similar value to the original model. However, in contrast to the identical model, which follows the mass evolution of the original model exactly, the perturbed models initially lose very less mass in comparison. This phase of low mass-loss rate is qualitatively similar to the original model during its own infancy, i.e., during $t=0-1\,\gyr$. This is due to the relative compactness of a King profile compared to the final equilibrium profile. This can be seen in the top panel of \autoref{fig:pert}, where the $90\%\ \rl$s of both perturbed models are a few times smaller than the original model at $\tpert$. In addition, the central dynamics of King profiles is also weaker than the equilibrium profile, due to a smaller central density. Again, this can be inferred from the $1\%\ \rl$s. Both of these attributes are expected to weaken the mass-loss \citep[e.g.,][]{Lamers2010}. However, as soon as the models converge to the equilibrium profile, the mass-loss rates become similar as well. These results demonstrate that the equilibrium profile is stable, i.e., any perturbations in the density profile diminish as a result of dynamical evolution. They also demonstrate that the shape of the equilibrium profile is not dependent on the evolutionary history of the model clusters but only on their current properties.

\section{Discussion}
\label{sec:discuss}

To understand the structure of GCs before core-collapse, \citet{Madsen_1996} modeled them as isothermal spheres. However, unlike the classical isothermal sphere, Madsen's version has a finite mass. This is achieved by forcing the phase space macro-states to be occupied only by an integral number of discrete particles \citep[][]{Simons_1994}. Where the classical isothermal sphere allows fractional particles to populate energy levels all the way to $r\to\infty$, forcing integral occupation results in a maximum energy level ($E_{\rm{max}}<0$) that can be occupied by a particle to be at a finite $r_{\rm{max}}$. Note that this cutoff energy is achieved in King models by an ad-hoc lowering of the distribution function, whereas in the Simons-Madsen formalism, this is a natural consequence of the physically well-motivated assumption of discrete particles. Using the distribution function that maximizes entropy for Maxwell-Boltzmann statistics \citep[][]{Simons_1994},

\begin{equation}
\label{eq:madsen}
    F_p = [[A\nu_p\exp{(-\beta E_p)}]]
\end{equation}
(where $[[.]]$ represents rounding down to the nearest integer, $F_p$ and $\nu_p$ represent the number of particles and micro-states with energy $E_p$, and $A$ and $\beta$ are constants), \citet{Madsen_1996} derived density profiles and showed that they are very similar to King profiles (see their Figure 1).

\citet{Barnes_2012} further developed this formalism and derived a smooth distribution function unlike the step-like \autoref{eq:madsen}. They incorporated discreteness effects by replacing the standard choice of the Stirling approximation ($\ln x! = x \ln x-x$), which is valid in the large-$x$ limit, with a better approximation in the derivation of their distribution function that maximizes entropy for Maxwell-Boltzmann statistics. We reproduce their equation for the distribution function below-

\begin{equation}
\label{eq:barnes}
    \ln (F+1) - \frac{264F+276}{576(F+1)(F+600/576)}+\mu + \beta \epsilon = 0.
\end{equation}
Here, $F=\nu f/\eta$. $f$ is the coarse-grained distribution function in ($\boldsymbol{x},\boldsymbol{v}$) phase-space, $\eta$ is the density of a phase-space element, and $\nu$ is the number of micro-states in a macro-state. $\epsilon=v^2/2+\Phi$ is the specific energy of a phase-space element, and $\mu$ and $\beta$ are the undetermined multipliers for the constraints of mass and energy conservation, respectively. 

The resulting density profiles match King profiles even better \citep[Figure 8;][]{Barnes_2012}. Thus, solving for the maximum entropy state of a system described by Maxwell-Boltzmann statistics naturally leads to King density profiles. To arrive at a specific density profile (e.g., a specific value of $\wk$) starting from \autoref{eq:barnes}, two assumptions that are degenerate with each other, need to be made. The first is the ratio of $f$ at the center and $\eta$, i.e., $f_c/\eta$. The second being the value of $\nu$. 

This may explain our result of the existence of an equilibrium density profile. \citet{Madsen_1996} argues that Maxwell-Boltzmann statistics should be appropriate to describe GCs since two-body relaxation plays a significant role in their evolution. Indeed, one of the core assumptions of our Monte Carlo models is that the evolution of GCs is dominated by two-body relaxation. Furthermore, \citet{Madsen_1996} also argues that even though a maximum entropy state does not exist for self-gravitating systems, GCs during the quasi-static evolution before core-collapse could be expected to resemble the maximum entropy state for a fixed potential. Since the values of $f_c/\eta$ and $\nu$ after a few $\trhm$ for our models should depend on the properties of stars/BHs and how they occupy the phase space, not on the initial density profile, and since $f_c/\eta$ and $\nu$ together uniquely specify a King density profile which is the maximum entropy state for a Maxwell-Boltzmann system, after a few $\trhm$, our models with different $\wk$ should all converge towards this specific profile.

\begin{figure}
    \centering
    \includegraphics[width=1\linewidth]{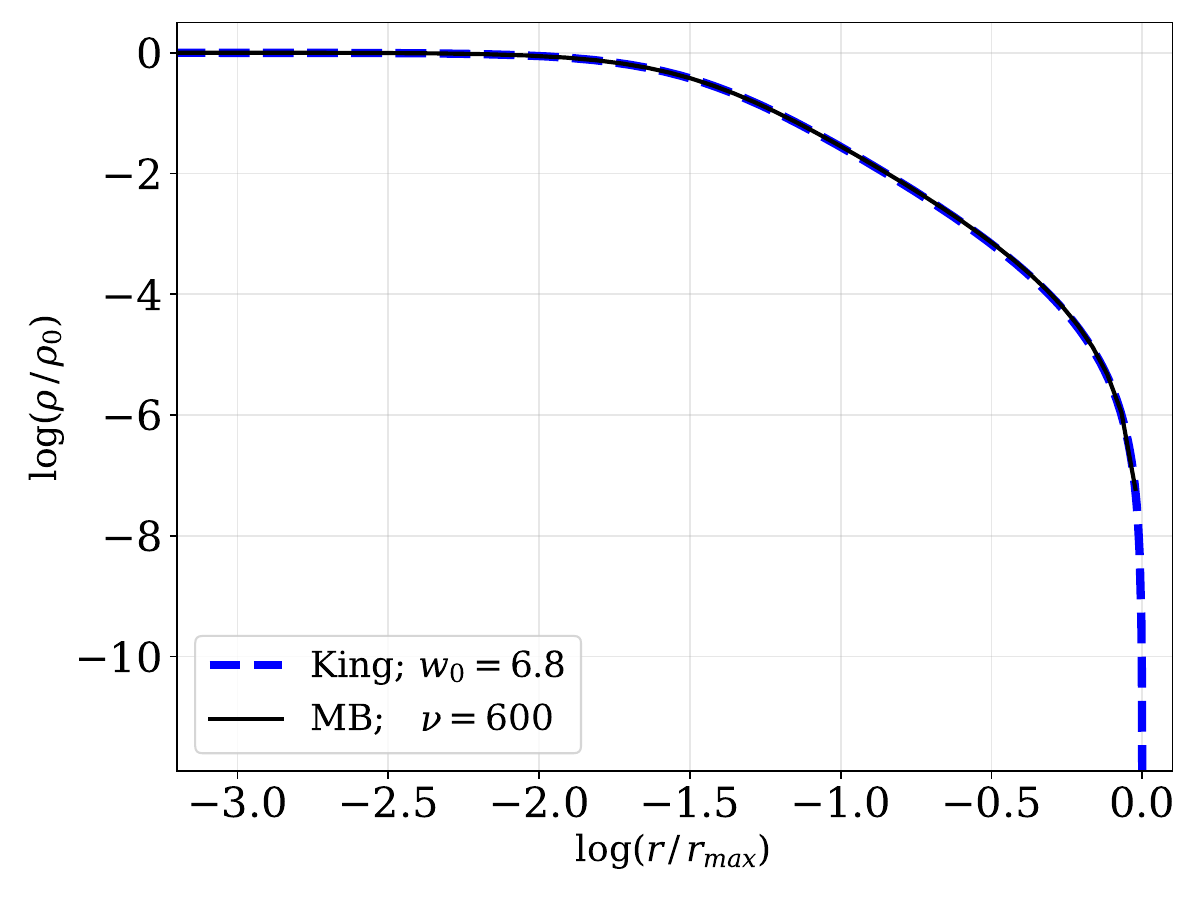}
    \caption{Normalized density profile derived using \autoref{eq:barnes} with $\nu=600$ and $f_c/\eta=1$ (black solid) compared to the King profile with $\wk=6.8$ (blue dashed).}
    \label{fig:barnes}
\end{figure}

To explore this further, in \autoref{fig:barnes}, we create an additional panel to complement Figure 8 of \citet{Barnes_2012}. Here, we compare the equilibrium King profile ($\wk=6.8$) we observe in our models with a profile derived using \autoref{eq:barnes} with $\nu=600$ and $f_c/\eta=1$. The similarity of the two profiles in \autoref{fig:barnes} suggests that describing our GC models in the Madsen-Barnes-Williams formalism requires $\nu\sim600$ if $f_c/\eta=1$. The latter condition means that the distribution function at the center is equal to the density of one phase-space element, and the former means that the number of micro-states in a macro-state is $\sim600$.

\section{Conclusions}
\label{sec:summ}

In this paper, we report our finding that there exists an equilibrium profile that massive and collisional star cluster models, characterized by an initial mass and a binding energy, tend to converge to, during the BH-burning phase, before core collapse. We checked this convergence for cluster models with vastly different King and EFF profiles (\autoref{fig:kingw0} and \autoref{fig:eff}). We also checked that this result is robust against variations in the initial binary fraction, $\fb$ (\autoref{fig:fb}). The key findings are as follows:

\begin{itemize}
    \item The projected number density profiles during the equilibrium phase are also almost identical for models, irrespective of their initial profile (\autoref{fig:numden}). The mock-observed surface brightness profiles also match for these models during the equilibrium phase (\autoref{fig:magV}).
    \item A set of relatively compact models with $\rv=0.5\,\pc$ also exhibit an equilibrium profile, before ejecting almost all of their BHs and going through core collapse within a Hubble time (\autoref{fig:rv05}). Interestingly, both equilibrium profiles for the model sets with $\rv=1$ and $0.5\,\pc$ are well matched by a single King model with $\wk=6.8$, scaled by their respective $\rv$s (\autoref{fig:fprof}). \citet{Lamers2010} also note this effect with a similar $\wk=7$. The final virial radius of the converged profile is about 7 times larger than the initial value.
    \item We also analyze the dynamical stability of the equilibrium profile using purely dynamical models comprising equal-mass point particles (\autoref{sec:exp}). To do this, we perturb the structure of a model during the equilibrium phase and show that under natural dynamical evolution, these perturbations decay and the models evolve back towards the equilibrium profile (\autoref{fig:dyn1}-\autoref{fig:pert}).
    \item King density profiles have been shown to naturally follow from the maximum entropy state in phase-space for a system described by Maxwell-Boltzmann statistics \citep[][\autoref{fig:barnes}]{Madsen_1996,Barnes_2012}. Assuming that GCs before core-collapse are well suited under this formalism \citep{Madsen_1996}, their final density profile should be uniquely defined by the number of micro-states in a macro-state and the ratio of the distribution function at the center and the density of a phase-space element. After a few $\trhm$, these parameters should not depend on the initial distribution. This may explain the evolution of GC models towards an equilibrium profile as a consequence of the tendency to maximize entropy.
\end{itemize}

On the one hand, this result suggests that, by observing the present day density profile of a GC, it should be impossible to infer its initial density profile, as the initial profile is degenerate in that context. However, it may be possible to break this degeneracy with some other observables. E.g., \citet{Dickson_2026} show that it is possible to infer initial GC conditions, including the half-mass radius/density, using multi-dimensional present-day observations. But, notice that $\rhm$ is \emph{not} a good indicator of the density profile or the King $\wk$. This is evident in \autoref{fig:kingw0}, i.e., the $\rhm$s of all the models are very similar at $t=0$ and remain so throughout the evolution. What $\rhm$ does constrain, is the initial cluster size, characterized by $\rv$ in our study. On the other hand, this result also means that the initial density profile is a parameter which provides freedom to modelers to be able to alter the evolutionary history of a model cluster, e.g., its BH retention history and possibly its binary BH merger history, without changing the final profile. In any case, we believe that this is an important and interesting result for the field of stellar dynamics, particularly in the context of GC modeling.

\begin{acknowledgments}
We acknowledge the support from the start-up funding of Zhejiang University and Zhejiang provincial top level research support program. The simulations and analysis presented in this article were carried out on the SilkRiver Supercomputer of Zhejiang University.
\end{acknowledgments}


\begin{thebibliography}{}
\expandafter\ifx\csname natexlab\endcsname\relax\def\natexlab#1{#1}\fi
\providecommand{\url}[1]{\href{#1}{#1}}
\providecommand{\dodoi}[1]{doi:~\href{http://doi.org/#1}{\nolinkurl{#1}}}
\providecommand{\doeprint}[1]{\href{http://ascl.net/#1}{\nolinkurl{http://ascl.net/#1}}}
\providecommand{\doarXiv}[1]{\href{https://arxiv.org/abs/#1}{\nolinkurl{https://arxiv.org/abs/#1}}}

\bibitem[{H.~A. {Abt}(1983){Abt}}]{Abt_1983}
{Abt}, H.~A. 1983, \bibinfo{title}{{Normal and abnormal binary frequencies.},} \araa, 21, 343, \dodoi{10.1146/annurev.aa.21.090183.002015}

\bibitem[{M. {Arca Sedda} {et~al.}(2026){Arca Sedda}, {Paiella}, {Ugolini}, {Santoliquido}, {Mestichelli}, {Usai}, {Simonato}, \& {Branchesi}}]{Sedda_2026}
{Arca Sedda}, M., {Paiella}, L., {Ugolini}, C., {et~al.} 2026, \bibinfo{title}{{Isolated or Dynamical? Tracing Black Hole Binary Formation through the Population of Gravitational-Wave Sources},} arXiv e-prints, arXiv:2603.20430, \dodoi{10.48550/arXiv.2603.20430}

\bibitem[{E.~I. {Barnes} \& L.~L.~R. {Williams}(2012){Barnes} \& {Williams}}]{Barnes_2012}
{Barnes}, E.~I., \& {Williams}, L. L.~R. 2012, \bibinfo{title}{{Entropy Production in Collisionless Systems. II. Arbitrary Phase-space Occupation Numbers},} \apj, 748, 144, \dodoi{10.1088/0004-637X/748/2/144}

\bibitem[{H. {Baumgardt} {et~al.}(2002){Baumgardt}, {Hut}, \& {Heggie}}]{Baumgardt_2002}
{Baumgardt}, H., {Hut}, P., \& {Heggie}, D.~C. 2002, \bibinfo{title}{{Long-term evolution of isolated N-body systems},} \mnras, 336, 1069, \dodoi{10.1046/j.1365-8711.2002.05736.x}

\bibitem[{K. {Breivik} {et~al.}(2020){Breivik}, {Coughlin}, {Zevin}, {Rodriguez}, {Kremer}, {Ye}, {Andrews}, {Kurkowski}, {Digman}, {Larson}, \& {Rasio}}]{Breivik_2020}
{Breivik}, K., {Coughlin}, S., {Zevin}, M., {et~al.} 2020, \bibinfo{title}{{COSMIC Variance in Binary Population Synthesis},} \apj, 898, 71, \dodoi{10.3847/1538-4357/ab9d85}

\bibitem[{J.~P. {Brodie} \& J. {Strader}(2006){Brodie} \& {Strader}}]{Brodie_2006}
{Brodie}, J.~P., \& {Strader}, J. 2006, \bibinfo{title}{{Extragalactic Globular Clusters and Galaxy Formation},} \araa, 44, 193, \dodoi{10.1146/annurev.astro.44.051905.092441}

\bibitem[{S. Chatterjee {et~al.}(2010)Chatterjee, Fregeau, Umbreit, \& Rasio}]{Chatterjee2010}
Chatterjee, S., Fregeau, J.~M., Umbreit, S., \& Rasio, F.~A. 2010, \bibinfo{title}{Monte Carlo simulations of globular cluster evolution. V. Binary stellar evolution,} Astrophysical Journal, 719, 915, \dodoi{10.1088/0004-637X/719/1/915}

\bibitem[{S. Chatterjee {et~al.}(2013)Chatterjee, Umbreit, Fregeau, \& Rasio}]{Chatterjee2013}
Chatterjee, S., Umbreit, S., Fregeau, J.~M., \& Rasio, F.~A. 2013, \bibinfo{title}{Understanding the dynamical state of globular clusters: Core-collapsed versus non-core-collapsed,} Monthly Notices of the Royal Astronomical Society, 429, 2881, \dodoi{10.1093/mnras/sts464}

\bibitem[{H. {Cohn}(1980){Cohn}}]{Cohn_1980}
{Cohn}, H. 1980, \bibinfo{title}{{Late core collapse in star clusters and the gravothermal instability},} \apj, 242, 765, \dodoi{10.1086/158511}

\bibitem[{T.~J.~L. {de Boer} {et~al.}(2019){de Boer}, {Gieles}, {Balbinot}, {H{\'e}nault-Brunet}, {Sollima}, {Watkins}, \& {Claydon}}]{DeBoer_2019}
{de Boer}, T.~J.~L., {Gieles}, M., {Balbinot}, E., {et~al.} 2019, \bibinfo{title}{{Globular cluster number density profiles using Gaia DR2},} \mnras, 485, 4906, \dodoi{10.1093/mnras/stz651}

\bibitem[{N. {Dickson} {et~al.}(2026){Dickson}, {H{\'e}nault-Brunet}, {Fronimos Pouliasis}, {Gieles}, \& {Smith}}]{Dickson_2026}
{Dickson}, N., {H{\'e}nault-Brunet}, V., {Fronimos Pouliasis}, F., {Gieles}, M., \& {Smith}, P.~J. 2026, \bibinfo{title}{{Fast Dynamical Modelling of Milky Way Globular Clusters -- I. Implications for Initial Cluster Densities},} arXiv e-prints, arXiv:2607.07010, \dodoi{10.48550/arXiv.2607.07010}

\bibitem[{K. El-Badry {et~al.}(2019)El-Badry, Quataert, Weisz, Choksi, \& Boylan-Kolchin}]{El-Badry2019a}
El-Badry, K., Quataert, E., Weisz, D.~R., Choksi, N., \& Boylan-Kolchin, M. 2019, \bibinfo{title}{The formation and hierarchical assembly of globular cluster populations,} Monthly Notices of the Royal Astronomical Society, 482, 4528, \dodoi{10.1093/mnras/sty3007}

\bibitem[{R.~A.~W. {Elson} {et~al.}(1987){Elson}, {Fall}, \& {Freeman}}]{Elson_1987}
{Elson}, R. A.~W., {Fall}, S.~M., \& {Freeman}, K.~C. 1987, \bibinfo{title}{{The Structure of Young Star Clusters in the Large Magellanic Cloud},} \apj, 323, 54, \dodoi{10.1086/165807}

\bibitem[{J.~M. Fregeau \& F.~A. Rasio(2007)Fregeau \& Rasio}]{Fregeau2007}
Fregeau, J.~M., \& Rasio, F.~A. 2007, \bibinfo{title}{Monte Carlo Simulations of Globular Cluster Evolution. IV. Direct Integration of Strong Interactions,} The Astrophysical Journal, 658, 1047, \dodoi{10.1086/511809}

\bibitem[{M. Gieles {et~al.}(2011)Gieles, Heggie, \& Zhao}]{Gieles2011}
Gieles, M., Heggie, D.~C., \& Zhao, H. 2011, \bibinfo{title}{The life cycle of star clusters in a tidal field,} Monthly Notices of the Royal Astronomical Society, 413, 2509, \dodoi{10.1111/j.1365-2966.2011.18320.x}

\bibitem[{M. {Giersz} \& D.~C. {Heggie}(1996){Giersz} \& {Heggie}}]{Giersz_1996}
{Giersz}, M., \& {Heggie}, D.~C. 1996, \bibinfo{title}{{Statistics of N-body simulations - III. Unequal masses},} \mnras, 279, 1037, \dodoi{10.1093/mnras/279.3.1037}

\bibitem[{S. {Goswami} {et~al.}(2012){Goswami}, {Umbreit}, {Bierbaum}, \& {Rasio}}]{Goswami_2012}
{Goswami}, S., {Umbreit}, S., {Bierbaum}, M., \& {Rasio}, F.~A. 2012, \bibinfo{title}{{Formation of Massive Black Holes in Dense Star Clusters. II. Initial Mass Function and Primordial Mass Segregation},} \apj, 752, 43, \dodoi{10.1088/0004-637X/752/1/43}

\bibitem[{D. {Heggie} \& P. {Hut}(2003){Heggie} \& {Hut}}]{HeggieHut_2003}
{Heggie}, D., \& {Hut}, P. 2003, {The Gravitational Million-Body Problem: A Multidisciplinary Approach to Star Cluster Dynamics}

\bibitem[{M. {H{\'e}non}(1961){H{\'e}non}}]{Henon_1961}
{H{\'e}non}, M. 1961, \bibinfo{title}{{Sur l'{\'e}volution dynamique des amas globulaires},} Annales d'Astrophysique, 24, 369

\bibitem[{M. {H{\'e}non}(1965){H{\'e}non}}]{Henon_1965}
{H{\'e}non}, M. 1965, \bibinfo{title}{{Sur l'{\'e}volution dynamique des amas globulaires. II. Amas isol\&eacute},} Annales d'Astrophysique, 28, 62

\bibitem[{M. {H{\'e}non}(1971{\natexlab{a}}){H{\'e}non}}]{Henon_1971a}
{H{\'e}non}, M. 1971{\natexlab{a}}, \bibinfo{title}{{Monte Carlo Models of Star Clusters (Part of the Proceedings of the IAU Colloquium No. 10, held in Cambridge, England, August 12-15, 1970.)},} \apss, 13, 284, \dodoi{10.1007/BF00649159}

\bibitem[{M. {H{\'e}non}(1971{\natexlab{b}}){H{\'e}non}}]{Henon_1971b}
{H{\'e}non}, M. 1971{\natexlab{b}}, \bibinfo{title}{{The Monte Carlo Method (Papers appear in the Proceedings of IAU Colloquium No. 10 Gravitational N-Body Problem (ed. by Myron Lecar), R. Reidel Publ. Co. , Dordrecht-Holland.)},} \apss, 14, 151, \dodoi{10.1007/BF00649201}

\bibitem[{M. {H{\'e}non}(1975){H{\'e}non}}]{Henon_1975}
{H{\'e}non}, M. 1975, \bibinfo{title}{{Two Recent Developments Concerning the Monte Carlo Method},} in IAU Symposium, Vol.~69, Dynamics of Stellar Systems, ed. A.~{Hayli}, 133

\bibitem[{M. {H{\'e}non}(2011{\natexlab{a}}){H{\'e}non}}]{Henon_2011}
{H{\'e}non}, M. 2011{\natexlab{a}}, \bibinfo{title}{{On the dynamical evolution of globular clusters},} arXiv e-prints, arXiv:1103.3499, \dodoi{10.48550/arXiv.1103.3499}

\bibitem[{M. {H{\'e}non}(2011{\natexlab{b}}){H{\'e}non}}]{Henon2_2011}
{H{\'e}non}, M. 2011{\natexlab{b}}, \bibinfo{title}{{On the dynamical evolution of globular clusters. II- The isolated cluster},} arXiv e-prints, arXiv:1103.3498, \dodoi{10.48550/arXiv.1103.3498}

\bibitem[{J. {Hjorth} \& L.~L.~R. {Williams}(2010){Hjorth} \& {Williams}}]{Hjorth_2010}
{Hjorth}, J., \& {Williams}, L. L.~R. 2010, \bibinfo{title}{{Statistical Mechanics of Collisionless Orbits. I. Origin of Central Cusps in Dark-matter Halos},} \apj, 722, 851, \dodoi{10.1088/0004-637X/722/1/851}

\bibitem[{J.~R. Hurley {et~al.}(2000)Hurley, Pols, \& Tout}]{Hurley2000}
Hurley, J.~R., Pols, O.~R., \& Tout, C.~A. 2000, \bibinfo{title}{Comprehensive analytic formulae for stellar evolution as a function of mass and metallicity,} Monthly Notices of the Royal Astronomical Society, 315, 543, \dodoi{10.1046/j.1365-8711.2000.03426.x}

\bibitem[{J.~R. Hurley {et~al.}(2002)Hurley, Tout, \& Pols}]{Hurley2002}
Hurley, J.~R., Tout, C.~A., \& Pols, O.~R. 2002, \bibinfo{title}{Evolution of binary stars and the effect of tides on binary populations,} Monthly Notices of the Royal Astronomical Society, 329, 897, \dodoi{10.1046/j.1365-8711.2002.05038.x}

\bibitem[{J.~H. {Jeans}(1919){Jeans}}]{Jeans_1919}
{Jeans}, J.~H. 1919, \bibinfo{title}{{The origin of binary systems},} \mnras, 79, 408, \dodoi{10.1093/mnras/79.6.408}

\bibitem[{K.~J. {Joshi} {et~al.}(2001){Joshi}, {Nave}, \& {Rasio}}]{Joshi_2001}
{Joshi}, K.~J., {Nave}, C.~P., \& {Rasio}, F.~A. 2001, \bibinfo{title}{{Monte Carlo Simulations of Globular Cluster Evolution. II. Mass Spectra, Stellar Evolution, and Lifetimes in the Galaxy},} \apj, 550, 691, \dodoi{10.1086/319771}

\bibitem[{K.~J. {Joshi} {et~al.}(2000){Joshi}, {Rasio}, \& {Portegies Zwart}}]{Joshi_2000}
{Joshi}, K.~J., {Rasio}, F.~A., \& {Portegies Zwart}, S. 2000, \bibinfo{title}{{Monte Carlo Simulations of Globular Cluster Evolution. I. Method and Test Calculations},} \apj, 540, 969, \dodoi{10.1086/309350}

\bibitem[{A. {Khurana} \& S. {Chatterjee}(2025){Khurana} \& {Chatterjee}}]{Khurana_2025}
{Khurana}, A., \& {Chatterjee}, S. 2025, \bibinfo{title}{{The Role of High-mass Stellar Binaries in the Formation of High-mass Black Holes in Dense Star Clusters},} \apj, 986, 82, \dodoi{10.3847/1538-4357/add337}

\bibitem[{A. {Khurana} {et~al.}(2023){Khurana}, {Chawla}, \& {Chatterjee}}]{Khurana_2023}
{Khurana}, A., {Chawla}, C., \& {Chatterjee}, S. 2023, \bibinfo{title}{{Dynamically Forming Extremely Low-mass White Dwarf Binaries in Wide Orbits},} \apj, 949, 102, \dodoi{10.3847/1538-4357/acc8d6}

\bibitem[{I.~R. {King}(1966){King}}]{King_1966}
{King}, I.~R. 1966, \bibinfo{title}{{The structure of star clusters. III. Some simple dynamical models},} \aj, 71, 64, \dodoi{10.1086/109857}

\bibitem[{K. {Kremer} {et~al.}(2020){Kremer}, {Ye}, {Chatterjee}, {Rodriguez}, \& {Rasio}}]{Kremer_2020}
{Kremer}, K., {Ye}, C.~S., {Chatterjee}, S., {Rodriguez}, C.~L., \& {Rasio}, F.~A. 2020, \bibinfo{title}{{The Role of ``black hole burning'' in the evolution of dense star clusters},} in IAU Symposium, Vol. 351, Star Clusters: From the Milky Way to the Early Universe, ed. A.~{Bragaglia}, M.~{Davies}, A.~{Sills}, \& E.~{Vesperini}, 357--366, \dodoi{10.1017/S1743921319007269}

\bibitem[{P. Kroupa(2001)Kroupa}]{Kroupa2001}
Kroupa, P. 2001, \bibinfo{title}{On the variation of the initial mass function,} Monthly Notices of the Royal Astronomical Society, 322, 231, \dodoi{10.1046/j.1365-8711.2001.04022.x}

\bibitem[{J.~M. Kruijssen(2015)Kruijssen}]{Kruijssen2015}
Kruijssen, J.~M. 2015, \bibinfo{title}{Globular clusters as the relics of regular star formation in 'normal' high-redshift galaxies,} Monthly Notices of the Royal Astronomical Society, 454, 1658, \dodoi{10.1093/mnras/stv2026}

\bibitem[{S.~R. {Kulkarni} {et~al.}(1993){Kulkarni}, {Hut}, \& {McMillan}}]{Kulkarni_1993}
{Kulkarni}, S.~R., {Hut}, P., \& {McMillan}, S. 1993, \bibinfo{title}{{Stellar black holes in globular clusters},} \nat, 364, 421, \dodoi{10.1038/364421a0}

\bibitem[{A.~H.~W. {K{\"u}pper} {et~al.}(2008){K{\"u}pper}, {Kroupa}, \& {Baumgardt}}]{Kupper_2008}
{K{\"u}pper}, A. H.~W., {Kroupa}, P., \& {Baumgardt}, H. 2008, \bibinfo{title}{{The main sequence of star clusters},} \mnras, 389, 889, \dodoi{10.1111/j.1365-2966.2008.13598.x}

\bibitem[{H.~J. Lamers {et~al.}(2010)Lamers, Baumgardt, \& Gieles}]{Lamers2010}
Lamers, H.~J., Baumgardt, H., \& Gieles, M. 2010, \bibinfo{title}{Mass-loss rates and the mass evolution of star clusters,} Monthly Notices of the Royal Astronomical Society, 409, 305, \dodoi{10.1111/j.1365-2966.2010.17309.x}

\bibitem[{J. {Liang} {et~al.}(2024){Liang}, {Jiang}, {Danieli}, {Benson}, \& {Hopkins}}]{Liang_2024}
{Liang}, J., {Jiang}, F., {Danieli}, S., {Benson}, A., \& {Hopkins}, P. 2024, \bibinfo{title}{{Constrain the Dark-matter Distribution of Ultra-diffuse Galaxies with Globular-cluster Mass Segregation: A Case Study with NGC5846-UDG1},} \apj, 964, 53, \dodoi{10.3847/1538-4357/ad1ee3}

\bibitem[{D. {Lynden-Bell} \& P.~P. {Eggleton}(1980){Lynden-Bell} \& {Eggleton}}]{Lynden-Bell_1980}
{Lynden-Bell}, D., \& {Eggleton}, P.~P. 1980, \bibinfo{title}{{On the consequences of the gravothermal catastrophe},} \mnras, 191, 483, \dodoi{10.1093/mnras/191.3.483}

\bibitem[{A.~D. MacKey {et~al.}(2008)MacKey, Wilkinson, Davies, \& Gilmore}]{Mackey2008}
MacKey, A.~D., Wilkinson, M.~I., Davies, M.~B., \& Gilmore, G.~F. 2008, \bibinfo{title}{Black holes and core expansion in massive star clusters,} Monthly Notices of the Royal Astronomical Society, 386, 65, \dodoi{10.1111/j.1365-2966.2008.13052.x}

\bibitem[{J. {Madsen}(1996){Madsen}}]{Madsen_1996}
{Madsen}, J. 1996, \bibinfo{title}{{Finite-mass isothermal spheres and the structure of globular clusters},} \mnras, 280, 1089, \dodoi{10.1093/mnras/280.4.1089}

\bibitem[{D. {Massari} {et~al.}(2023){Massari}, {Aguado-Agelet}, {Monelli}, {Cassisi}, {Pancino}, {Saracino}, {Gallart}, {Ruiz-Lara}, {Fern{\'a}ndez-Alvar}, {Surot}, \& et~al.}]{Massari_2023}
{Massari}, D., {Aguado-Agelet}, F., {Monelli}, M., {et~al.} 2023, \bibinfo{title}{{Cluster Ages to Reconstruct the Milky Way Assembly (CARMA). I. The final word on the origin of NGC 6388 and NGC 6441},} \aap, 680, A20, \dodoi{10.1051/0004-6361/202347289}

\bibitem[{D.~E. {McLaughlin} \& R.~P. {van der Marel}(2005){McLaughlin} \& {van der Marel}}]{McLaughlin_2005}
{McLaughlin}, D.~E., \& {van der Marel}, R.~P. 2005, \bibinfo{title}{{Resolved Massive Star Clusters in the Milky Way and Its Satellites: Brightness Profiles and a Catalog of Fundamental Parameters},} \apjs, 161, 304, \dodoi{10.1086/497429}

\bibitem[{G. {Meylan} \& D.~C. {Heggie}(1997){Meylan} \& {Heggie}}]{Meylan_1997}
{Meylan}, G., \& {Heggie}, D.~C. 1997, \bibinfo{title}{{Internal dynamics of globular clusters},} \aapr, 8, 1, \dodoi{10.1007/s001590050008}

\bibitem[{P. {Miocchi} {et~al.}(2013){Miocchi}, {Lanzoni}, {Ferraro}, {Dalessandro}, {Vesperini}, {Pasquato}, {Beccari}, {Pallanca}, \& {Sanna}}]{Miocchi_2013}
{Miocchi}, P., {Lanzoni}, B., {Ferraro}, F.~R., {et~al.} 2013, \bibinfo{title}{{Star Count Density Profiles and Structural Parameters of 26 Galactic Globular Clusters},} \apj, 774, 151, \dodoi{10.1088/0004-637X/774/2/151}

\bibitem[{M. Morscher {et~al.}(2015)Morscher, Pattabiraman, Rodriguez, Rasio, \& Umbreit}]{Morscher2015}
Morscher, M., Pattabiraman, B., Rodriguez, C., Rasio, F.~A., \& Umbreit, S. 2015, \bibinfo{title}{The dynamical evolution of stellar black holes in globular clusters,} Astrophysical Journal Letters, 800, \dodoi{10.1088/0004-637X/800/1/9}

\bibitem[{M. {Morscher} {et~al.}(2013){Morscher}, {Umbreit}, {Farr}, \& {Rasio}}]{Morscher_2013}
{Morscher}, M., {Umbreit}, S., {Farr}, W.~M., \& {Rasio}, F.~A. 2013, \bibinfo{title}{{Retention of Stellar-mass Black Holes in Globular Clusters},} \apjl, 763, L15, \dodoi{10.1088/2041-8205/763/1/L15}

\bibitem[{O. Newton {et~al.}(2025)Newton, Davies, Pfeffer, Crain, Kruijssen, Pontzen, \& Bastian}]{Newton_2025}
Newton, O., Davies, J.~J., Pfeffer, J., {et~al.} 2025, \bibinfo{title}{The formation and disruption of globular cluster populations in simulations of present-day L* galaxies with controlled assembly histories,} Monthly Notices of the Royal Astronomical Society, 542, 591, \dodoi{10.1093/mnras/staf1226}

\bibitem[{B. {Pattabiraman} {et~al.}(2013){Pattabiraman}, {Umbreit}, {Liao}, {Choudhary}, {Kalogera}, {Memik}, \& {Rasio}}]{Pattabiraman_2013}
{Pattabiraman}, B., {Umbreit}, S., {Liao}, W.-k., {et~al.} 2013, \bibinfo{title}{{A Parallel Monte Carlo Code for Simulating Collisional N-body Systems},} \apjs, 204, 15, \dodoi{10.1088/0067-0049/204/2/15}

\bibitem[{H.~C. {Plummer}(1911){Plummer}}]{Plummer_1911}
{Plummer}, H.~C. 1911, \bibinfo{title}{{On the problem of distribution in globular star clusters},} \mnras, 71, 460, \dodoi{10.1093/mnras/71.5.460}

\bibitem[{S.~F. {Portegies Zwart} {et~al.}(2010){Portegies Zwart}, {McMillan}, \& {Gieles}}]{PortegiesZwart_2010}
{Portegies Zwart}, S.~F., {McMillan}, S. L.~W., \& {Gieles}, M. 2010, \bibinfo{title}{{Young Massive Star Clusters},} \araa, 48, 431, \dodoi{10.1146/annurev-astro-081309-130834}

\bibitem[{M. Reina-Campos {et~al.}(2023)Reina-Campos, Trujillo-Gomez, Pfeffer, Sills, Deason, Crain, \& Kruijssen}]{Reina-Campos_2023}
Reina-Campos, M., Trujillo-Gomez, S., Pfeffer, J.~L., {et~al.} 2023, \bibinfo{title}{Constraining the shape of dark matter haloes with globular clusters and diffuse stellar light in the E-MOSAICS simulations,} Monthly Notices of the Royal Astronomical Society, 521, 6368, \dodoi{10.1093/mnras/stad920}

\bibitem[{C. {Rodrigo} \& E. {Solano}(2020){Rodrigo} \& {Solano}}]{Rodrigo_2020}
{Rodrigo}, C., \& {Solano}, E. 2020, \bibinfo{title}{{The SVO Filter Profile Service},} in XIV.0 Scientific Meeting (virtual) of the Spanish Astronomical Society, 182

\bibitem[{C.~L. Rodriguez {et~al.}(2015)Rodriguez, Morscher, Pattabiraman, Chatterjee, Haster, \& Rasio}]{Rodriguez2015}
Rodriguez, C.~L., Morscher, M., Pattabiraman, B., {et~al.} 2015, \bibinfo{title}{Binary Black Hole Mergers from Globular Clusters: Implications for Advanced LIGO,} Physical Review Letters, 115, 1, \dodoi{10.1103/PhysRevLett.115.051101}

\bibitem[{C.~L. {Rodriguez} {et~al.}(2016){Rodriguez}, {Morscher}, {Wang}, {Chatterjee}, {Rasio}, \& {Spurzem}}]{Rodriguez_2016}
{Rodriguez}, C.~L., {Morscher}, M., {Wang}, L., {et~al.} 2016, \bibinfo{title}{{Million-body star cluster simulations: comparisons between Monte Carlo and direct N-body},} \mnras, 463, 2109, \dodoi{10.1093/mnras/stw2121}

\bibitem[{C.~L. Rodriguez {et~al.}(2022)Rodriguez, Weatherford, Coughlin, Amaro-Seoane, Breivik, Chatterjee, Fragione, Kıroğlu, Kremer, Rui, Ye, Zevin, \& Rasio}]{Rodriguez2022}
Rodriguez, C.~L., Weatherford, N.~C., Coughlin, S.~C., {et~al.} 2022, \bibinfo{title}{Modeling Dense Star Clusters in the Milky Way and beyond with the Cluster Monte Carlo Code,} The Astrophysical Journal Supplement Series, 258, 22, \dodoi{10.3847/1538-4365/ac2edf}

\bibitem[{N.~Z. Rui {et~al.}(2021{\natexlab{a}})Rui, Kremer, Weatherford, Chatterjee, Rasio, Rodriguez, \& Ye}]{Rui2021}
Rui, N.~Z., Kremer, K., Weatherford, N.~C., {et~al.} 2021{\natexlab{a}}, \bibinfo{title}{Matching Globular Cluster Models to Observations,} The Astrophysical Journal, 912, 102, \dodoi{10.3847/1538-4357/abed49}

\bibitem[{N.~Z. Rui {et~al.}(2021{\natexlab{b}})Rui, Kremer, Weatherford, Chatterjee, Rasio, Rodriguez, \& Ye}]{rui_2021}
Rui, N.~Z., Kremer, K., Weatherford, N.~C., {et~al.} 2021{\natexlab{b}}, NicholasRui/cmctoolkit: First release, 1.0 Zenodo, \dodoi{10.5281/zenodo.4579951}

\bibitem[{K. {Sharma} \& C.~L. {Rodriguez}(2025){Sharma} \& {Rodriguez}}]{Sharma_2025}
{Sharma}, K., \& {Rodriguez}, C.~L. 2025, \bibinfo{title}{{From Young Massive Clusters to Old Globular Clusters: Density Profile Evolution and Intermediate-mass Black Hole Formation},} \apj, 983, 162, \dodoi{10.3847/1538-4357/adbbdf}

\bibitem[{S. {Simons}(1994){Simons}}]{Simons_1994}
{Simons}, S. 1994, \bibinfo{title}{{Statistical mechanical distributions for small numbers of systems},} American Journal of Physics, 62, 515, \dodoi{10.1119/1.17563}

\bibitem[{L. {Spitzer}(1969){Spitzer}}]{Spitzer_1969}
{Spitzer}, Jr., L. 1969, \bibinfo{title}{{Equipartition and the Formation of Compact Nuclei in Spherical Stellar Systems},} \apjl, 158, L139, \dodoi{10.1086/180451}

\bibitem[{M. Trenti {et~al.}(2010)Trenti, Vesperini, \& Pasquato}]{Trenti2010}
Trenti, M., Vesperini, E., \& Pasquato, M. 2010, \bibinfo{title}{Tidal disruption, global mass function, and structural parameter evolution in star clusters,} Astrophysical Journal, 708, 1598, \dodoi{10.1088/0004-637X/708/2/1598}

\bibitem[{L.~M. {Valenzuela}(2023){Valenzuela}}]{Valenzuela_2023}
{Valenzuela}, L.~M. 2023, \bibinfo{title}{{Lights in the Dark: Globular clusters as dark matter tracers},} in Memorie della Societa Astronomica Italiana, Vol.~94, 47, \dodoi{10.36116/MEMSAIT_94N3.2023.47}

\bibitem[{L.~M. {Valenzuela} {et~al.}(2024){Valenzuela}, {Remus}, {McKenzie}, \& {Forbes}}]{Valenzuela_2024}
{Valenzuela}, L.~M., {Remus}, R.-S., {McKenzie}, M., \& {Forbes}, D.~A. 2024, \bibinfo{title}{{Galaxy archaeology for wet mergers: Globular cluster age distributions in the Milky Way and nearby galaxies},} \aap, 687, A104, \dodoi{10.1051/0004-6361/202348010}

\bibitem[{N.~C. Weatherford {et~al.}(2021)Weatherford, Fragione, Kremer, Chatterjee, Ye, Rodriguez, \& Rasio}]{Weatherford2021}
Weatherford, N.~C., Fragione, G., Kremer, K., {et~al.} 2021, \bibinfo{title}{Black Hole Mergers from Star Clusters with Top-heavy Initial Mass Functions,} The Astrophysical Journal Letters, 907, L25, \dodoi{10.3847/2041-8213/abd79c}

\bibitem[{L.~L. Williams {et~al.}(2012)Williams, Barnes, \& Hjorth}]{Williams2012}
Williams, L.~L., Barnes, E.~I., \& Hjorth, J. 2012, \bibinfo{title}{Collisionless dynamics in globular clusters,} Monthly Notices of the Royal Astronomical Society, 423, 3589, \dodoi{10.1111/J.1365-2966.2012.21158.X}

\bibitem[{L.~L.~R. {Williams} \& J. {Hjorth}(2010){Williams} \& {Hjorth}}]{Williams_2010}
{Williams}, L. L.~R., \& {Hjorth}, J. 2010, \bibinfo{title}{{Statistical Mechanics of Collisionless Orbits. II. Structure of Halos},} \apj, 722, 856, \dodoi{10.1088/0004-637X/722/1/856}

\bibitem[{C.~P. {Wilson}(1975){Wilson}}]{Wilson_1975}
{Wilson}, C.~P. 1975, \bibinfo{title}{{Dynamical models of elliptical galaxies.},} \aj, 80, 175, \dodoi{10.1086/111729}

\end{thebibliography}
\bibliographystyle{aasjournalv7}

\end{document}